\pdfoutput=1

\documentclass[a4paper,12pt,aps,superscriptaddress]{article}
\usepackage{amsmath}

\usepackage{amsfonts}
\usepackage{amssymb}

\usepackage{float}
\usepackage{epsfig}

\gdef\SM{S.V.~Mashkevich}
\gdef\JM{J.~Myrheim}
\gdef\KO{K.~Olaussen}
\gdef\SO{S.~Ouvry}

\newcommand{\be}{\begin{equation}}
\newcommand{\ee}{\end{equation}}
\newcommand{\bea}{\begin{eqnarray}}
\newcommand{\eea}{\end{eqnarray}}

\begin{document}
\title{Anyons and lowest Landau level Anyons}
\author{St\'ephane Ouvry$^{1,2}$\\
$^{1}$ Universit\'e  Paris-Sud, Laboratoire de Physique Th\'eorique et Mod\`eles
Statistiques\\ 
  UMR 8626, 91405 Orsay 
Cedex, France \\
$^{2}$ CNRS, Laboratoire de Physique Th\'eorique et Mod\`eles
Statistiques\\ 
  UMR 8626, 91405 Orsay 
Cedex, France \\}
\maketitle

\begin{abstract}
 Intermediate statistics  interpolating from Bose statistics to Fermi statistics are allowed  in two dimensions. This is due to 
 the particular  topology of
the two dimensional configuration space of identical particles, leading to non trivial braiding of particles around each other. One arrives at quantum many-body  states with a multivalued phase factor, which encodes the anyonic nature of particle windings. Bosons and fermions appear as two  limiting cases. Gauging away  the phase  leads to the so-called anyon model, where the charge of each particle interacts ''\`a la Aharonov-Bohm'' with the fluxes carried by the other particles.  The multivaluedness of the wave function has been traded off for topological interactions between ordinary particles. An alternative  Lagrangian formulation uses a topological Chern-Simons term in 2+1 dimensions. Taking  into account the short distance repulsion between particles leads to an Hamiltonian well  defined in perturbation theory,  where all perturbative divergences have disappeared. Together with numerical and semi-classical studies, perturbation theory is a basic analytical tool at disposal to study the model, since finding the exact $N$-body spectrum seems out of reach (except in the $2$-body case which is solvable, or for particular  classes of  $N$-body eigenstates which generalize some 2-body eigenstates). However, a simplification arises when the anyons are coupled to an external homogeneous magnetic field. In the case of a strong field, by projecting the system on its lowest Landau level (LLL, thus  the LLL-anyon model), the anyon model becomes solvable, i.e. the classes of exact eigenstates alluded to above provide for a complete interpolation from the LLL-Bose spectrum to the LLL-Fermi spectrum. Being a solvable model allows for an explicit knowledge of the equation of state and of the mean occupation number in the LLL, which do  interpolate from the Bose to the  Fermi cases. It  also provides for a  combinatorial interpretation of LLL-anyon braiding statistics in terms of occupation of single particle states. The LLL-anyon  model might also be relevant  experimentally: a gas of electrons in a strong magnetic field is known to exhibit a quantized Hall conductance, leading to the integer
and fractional quantum Hall effects.  Haldane/exclusion statistics, introduced  to describe FQHE edge excitations, is a priori different from anyon statistics, since it is not defined by  braiding considerations, but rather by  counting arguments in the space of available states. However, it has been shown to lead to the same kind of thermodynamics as the LLL-anyon thermodynamics (or, in other words, the LLL-anyon model is a microscopic quantum mechanical realization of Haldane's statistics).   The one dimensional Calogero model is also shown to have the same kind of thermodynamics as the LLL-anyons thermodynamics. This is not a coincidence: the LLL-anyon model and the Calogero model are intimately related, the latter being a particular limit of the former. Finally,  on the purely combinatorial side, the minimal difference partition problem -partition of integers with minimal difference constraints on their parts- can also be mapped on an abstract exclusion statistics  model with a  constant one-body density of states, which is neither the  LLL-anyon model nor the Calogero model.

\medskip\noindent {}

\end{abstract}

\vskip 0.2cm

{\bf Contributed paper to the IHP Poincar\'e Seminar 2007 ''Le Spin !''}

\vskip 0.2cm

\section{Introduction}

Quantum statistics, which is concerned with  quantum many-body wavefunctions of identical particles,   has a long history going back to  Bose and Fermi. The concept of statistics originates at the classical level in the Gibbs paradox, which is solved by means of the indiscernability postulate for identical particles. At the quantum level, the usual reasoning  shows that only two types of statistics can exist, bosonic or fermionic. Indeed, since 

\begin{itemize} 
\item  interchanging the positions of two identical particles can only amount to multiplying  their 2-body wavefunction by a phase factor,   
\item a double exchange puts back the particles at their original position, 
\item and  one usually  insists on the univaluedness of the  wavefunction, 
\end{itemize}
this phase factor can  be only 1 (boson) or -1 (fermion). 

However, non trivial phase factor should  be possible, since wavefunctions are anyway defined up to a phase. The configuration space of two, or more generally, $N$ identical particles  has to be  defined   cautiously    \cite{LM}: denoting by $C$ the configuration space of a single particle ($C=R^2$ for particles in the two-dimensional plane, $d=2$), the configuration space of $N$ particles should  be of the type $C^N/S_N$, where $C\times C\times...\times C=C^N$ and $S_N$ is the permutation group for $N$ identical particles. Quotienting by $S_N$  takes into account the identity of the particles which implies that  one cannot distinguish between two configurations related by an operation of the permutation group. One should also  subtract from $C^N$ the diagonal of the configuration space $D_N$, i.e. any configurations where two or more particles coincide. The reason  is,  having in mind  Fermi statistics, that the Pauli exclusion principle should be enforced in some way. A more precise argument is to have a valid classification of  paths in the $N$-particle configuration space, which would be ambiguous if two or more particles coinciding at some time is allowed (since they are identical,  did they  cross each other, or did they scatter off each other ?). It follows that  the configuration space of $N$ identical particles should be

\begin{equation}
\tilde C_N={C^N-D_N\over S_N}
\label{}
\end{equation}

Note that on this configuration space, a fermionic wavefunction is  multivalued  (two values 1 and -1), so there is no reason  not to allow  more general multivaluedness.  Here come some topological arguments,   which allow to distinguish between $d=2$ and $d>2$, and, as we will see later, which can be  related to spin considerations.  In 2 dimensions,   $ C^N$  is multiply connected and its  topology is non trivial: it is not possible to shrink  a path of a particle encircling another particle, due to the topological obstruction materialized by the latter. It follows that $\tilde C_N$ is multiply connected. This is not the case in dimensions higher than 2, where $C^N$  is simply connected, meaning that all paths made by a particle can be continuously deformed into each other, i.e. one cannot distinguish the interior from the exterior of a closed path of a particle around other particles. 

These arguments imply that the equivalent classes of paths (first homotopy group) in  $\tilde C_N$ are, when $d=2$,
in one-to-one correspondence with the elements of the braid group
\begin{equation}
\Pi_1(\tilde C_N)=B_N
\label{}
\end{equation}
whereas,  when $d>2$,
they are in one-to-one correspondence with the elements of the permutation group
\begin{equation}
\Pi_1(\tilde C_N)=S_N
\label{}
\end{equation}

The braid group generators  $\sigma_i$  interchange the position of particle $i$ with particle $i+1$.  This operation can be made in an anti-clockwise manner ($\sigma_i$) or a clockwise manner ($\sigma_i^{-1}$). Each braiding of $N$ particles consists of  a sequence of interchanges of pairs of neighboring particles  via the  $\sigma_i$'s and the $\sigma_i^{-1}$'s, with $i=1,2,...,N-1$. The braid group relations  list the equivalent braiding, i.e. braiding that can be continuously deformed  one into the other  without encountering a topological obstruction
\begin{equation}
 \sigma_i\sigma_{i+1}\sigma_i=\sigma_{i+1}\sigma_i\sigma_{i+1};
\quad \quad \sigma_i\sigma_j=\sigma_j\sigma_i \quad {\rm when} \quad |i-j|>2
\label{braid}
\end{equation}

\begin{figure}[htbp]
\epsfxsize=8cm
\centerline{\epsfbox{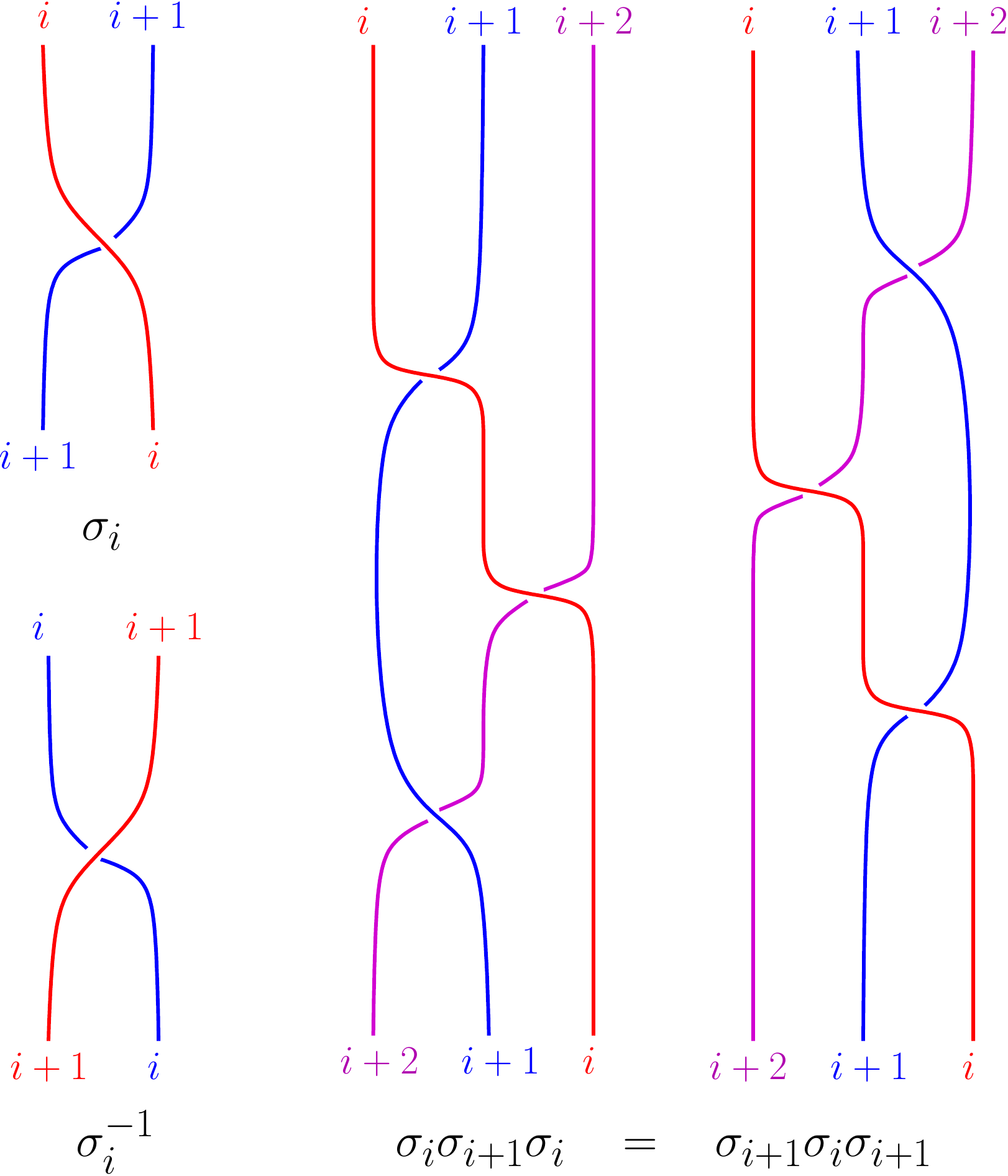}}
\caption{The braid group generators and their defining relations.}
\label{fig:braiding}
\end{figure}

Saying that  $d=2$ is different from  $d>2$  is nothing but recognizing that $\sigma_i\ne \sigma_i^{-1}$ when $d=2$, whereas $\sigma_i= \sigma_i^{-1}$ when $d>2$ ($\sigma_i$ can be continuously deformed into $\sigma_i^{-1}$  when  particles $i$ and $i+1$ are not stuck in a plane). It follows that when $d>2$,   the braid generators $\sigma_i$'s  defined by (\ref{braid}) with the additional constraint $\sigma_i= \sigma_i^{-1}$ are the permutation group generators.

Note also that the $d=2$ paradigm, $\sigma_i\ne \sigma_i^{-1}$, hints at an orientation  of the plane,  a hallmark of the presence of some sort of  magnetic field. This point will become apparent in the Aharonov-Bohm formulation of the anyon model. 

The fact that $C^N$ is multiply connected when $d=2$ and not when $d>2$ can also be related to  the rotation group  $O(d)$, and thus to some spin-statistics considerations \cite{Forte}.  When $d>2$, the rotation group is doubly connected, $[\Pi_1(O(d))=Z_2]$, its universal covering, for example when $d=3$, is $SU(2)$, which allows for either integer or half integer angular momentum states, that is to say either single valued or double valued representations of the rotation group.  On the other hand, when $d=2$, the rotation group is abelian and infinitely connected $[\Pi_1(O(2))=Z]$, its universal covering is the real line, that is to say  arbitrary angular momenta are possible, and therefore multivalued representations. One can see here a hint about the spin-statistics connection, where statistics and spin are trivial (Bose-Fermi statistics, integer-half integer spin) when $d>2$, and not when $d=2$.

Let us consider the simple one-dimensional irreducible representation of the braid group, which amounts to  a common phase factor $\exp(-i\pi\alpha)$ for each generator  $\sigma$ (and thus $\exp(i\pi\alpha)$ to $\sigma^{-1}$). It means that a non trivial phase has been associated with the winding of  particle $i$ around particle $i+1$. Higher dimensional representations (quantum vector states) are  possible -one speaks of non abelian anyons, in that case not only a non trivial  phase materializes  during a winding, but  also the direction  of the  vector state in the Hilbert space is affected- but they will not be discussed here (even though they might play a role in the discussion of certain FQHE fractions \cite{Nonab},  and, in a quite different perspective, in the definition of topologically protected fault-tolerant quantum computers \cite{Comp}).   

Clearly, when $d>2$, $\sigma_i= \sigma_i^{-1}$ implies $\alpha =0$ or $\alpha=1$, i.e. Bose or Fermi statistics (an interchange leaves the wavefunction unchanged or affected by a minus sign). 

From now on let us concentrate on $d=2$ and  denote the free many-body  wavefunction of $N$ identical particles by $\psi'(\vec r_1, \vec r_2,...,\vec r_N)$. Indeed, statistics should be  defined for free particles with Hamiltonian 
\begin{equation}\label{free}H'_N=\sum_{i=1}^N{\vec p_i^2\over 2m} \end{equation}
and special boundary conditions on the wavefunction, as in the Bose case (symmetric boundary condition) and the Fermi case (antisymmetric boundary condition).
As already said, $\psi'(\vec r_1,\vec r_2,...,\vec r_N)$ is affected by a phase $\exp(-i\pi\alpha)$  when particles $i$ and $i+1$ are interchanged: one can encode   this non trivial exchange property  by  defining
 \begin{equation}\label{def}
\psi'(\vec r_1,\vec r_2,...,\vec r_N)=\exp(-i\alpha\sum_{i< j}\theta_{ij})\psi(\vec r_1,\vec r_2,...,\vec r_N)
\end{equation}
where $\psi(\vec r_1,\vec r_2,...,\vec r_N)$ is a regular wavefunction, say bosonic by convention, and $\theta_{ij}$ is the angle between the vector  $\vec r_j-\vec r_i\equiv \vec r_{ij}$  and a fixed direction in the plane.  Indeed interchanging $i$ with $j$ amounts to $\theta_{ij}\to\theta_{ij}\pm \pi$, which altogether with the bosonic symmetry of $\psi(\vec r_1,\vec r_2,...,\vec r_N)$, leads to  
\begin{equation}
\psi'(\vec r_1,\vec r_2,...\vec r_j,...\vec r_i,...,\vec r_N)=\exp(\mp i\pi\alpha)\psi'(\vec r_1,\vec r_2,...\vec r_i,...\vec r_j,...,\vec r_N)
\end{equation}

By  the above bosonic convention for $\psi(\vec r_1,\vec r_2,...,\vec r_N)$, the statistical parameter $\alpha$ even (odd) integer corresponds to Bose (Fermi) statistics. It is defined modulo 2, since two  quanta of flux can always be gauged away by a regular gauge transformation while preserving the symmetry of the wavefunctions in the  Bose or Fermi systems. 
Indeed, (\ref{def}) can be interpreted  as a gauge transformation. Let us  compute the resulting Hamiltonian $H_N$ acting on  
$\psi(\vec r_1, \vec r_2,...,\vec r_N)$
\begin{equation}\label{defbis}
H_N=\sum_{i=1}^N{1\over 2m}(\vec p_i-\vec A(\vec r_i))^2 
\end{equation}
where 
\begin{equation}\label{A}
\vec A(\vec r_i)=\alpha\vec\partial_i(\sum_{k< l}\theta_{kl})=\alpha\sum_{j,j\ne i}{\vec k\wedge\vec r_{ij}\over r_{ij}^2}
\end{equation}
is the statistical potential vector associated with the multivalued phase (the gauge parameter).  The free multivalued wavefunction has been traded off for a regular bosonic wavefunction with  topological singular magnetic interactions. The statistical  potential vector (\ref{A}) can be viewed as the Aharonov-Bohm (A-B) potential vector that particle $i$ carrying a charge $e$ would feel due to the flux tube $\phi$ carried by the other particles, with $e$ and $\phi$ related to the statistical parameter $\alpha$  by $\alpha=e\phi/(2\pi)=\phi/\phi_0$  ($\phi_0=2\pi/e$ is the flux quantum in units $\hbar=1$). 
The resulting  composite charge-flux picture is known under the name of  anyon model \cite{W-82-0} since it describes particles with ''any'' (any-on)  statistics. 

\begin{figure}[htbp]
\epsfxsize=8cm
\centerline{\epsfbox{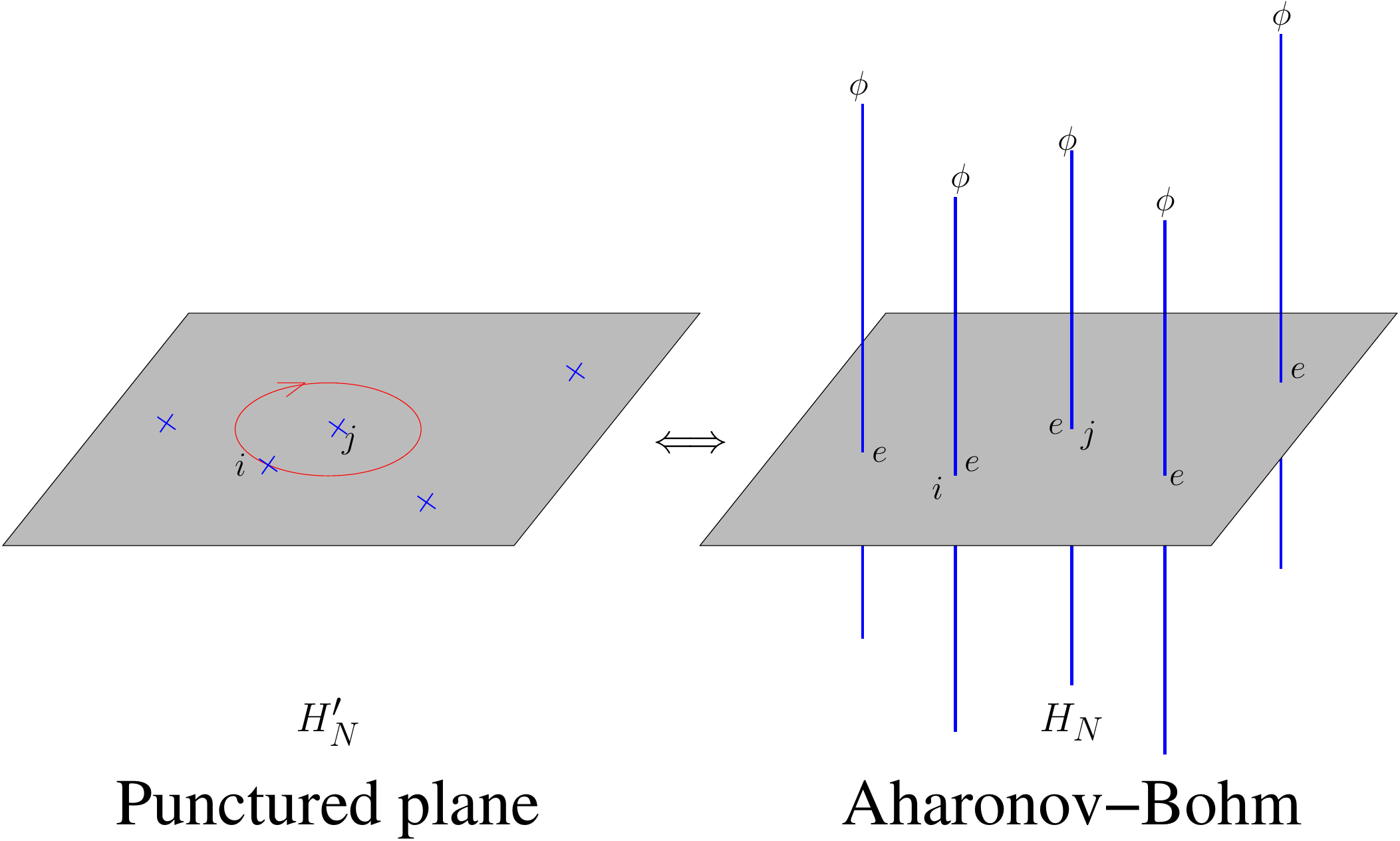}}
\caption{The two equivalent formulations of anyon statistics in terms, on the left, of a punctured plane and, on the right, of usual bosonic particles interacting via topological  A-B interactions. The loop of particle $i$ around particle $j$ cannot be continuously deformed to nothing due to the topological obstruction materialized by the puncture at the location of particle $j$. }
\label{fig:puncture}
\end{figure}

 Computing the field strength one obtains
\begin{equation}\label{mag}
 {\alpha\over e}\vec\partial_i\wedge\sum_{j, j\ne i}{\vec k\wedge\vec r_{ij}\over r_{ij}^2}={2\pi\alpha\over e}\sum_{j,j\ne i}\delta(\vec r_{ij})
\end{equation}
meaning that each particle carries an infinite singular magnetic field  with  flux $\phi=2\pi\alpha/e$.
The gauge transformation is singular since it does not preserve the field strength  (which vanishes  in the multivalued gauge and is  singular in the regular gauge). This is  due to the  singular behavior of the gauge parameter $\theta_{ij}$ when particle $i$ come close to particle $j$, thus the singular Dirac $\delta(\vec r_{ij})$ function in the field strength. 

It is not  surprising that topological A-B interactions are  at the heart of  quantum statistics. In its original form, the A-B effect \cite{AB} consists in the phase shift in electron interference due to the electromagnetic field, determined by the phase factor $\exp[(ie/\hbar c)\int_{\gamma}A_{\mu}dx^{\mu}]$ along a closed curve $\gamma$  passing through the beam along which the field strength vanishes. This effect\footnote{The effect was first experimentally confirmed by R. G. Chambers \cite{Cham}, then by A. Tonomura \cite{Tono}.} is  counter-intuitive to the usual understanding that the influence of a classical electromagnetic field  on a charged particle can only occur through the local action of the field strength. In the context of quantum statistics, it means that non trivial statistics  arise through  topological ''infinite''-distance interactions where no classical forces are present, as it should and as it is the case  for Bose and Fermi statistics. Finally,  singular magnetic fields  give an orientation to the plane, which, as already said, shows up in    $\sigma_i\ne \sigma_i^{-1}$.

All this can be equivalently restated in a Lagrange formulation which  describes again the system  in topological terms, i.e. free particles minimally coupled to a potential vector whose dynamics is not Coulomb-like (Maxwell Lagrangian) but rather Chern-Simons \cite{Chern}
 \begin{equation}
L_N=\sum_{i=1}^N ({1\over 2}m\vec v_i^2 + e(\vec A(\vec r_i)\vec v_i-A^0(\vec r_i)))-{\kappa\over 2}\epsilon_{\mu\nu\rho}\int d^2\vec r A^{\mu}\partial^{\nu}A^{\rho}
\end{equation}
with $\epsilon_{\mu\nu\rho}$ the completely antisymmetric tensor (the metric is  $(+,-,-), x^{\mu}=(t,\vec r )=(t,x,y), A_{\mu}=(A_o, A_x, A_y), \epsilon_{012}=\epsilon^{012}=+1)$.
Solving the Euler-Lagrange equations, in particular
 \begin{equation}
\partial^{\sigma} {\delta L_N\over \delta(\partial^{\sigma}A_0)}={\delta L_N\over \delta A^0}\to \kappa \vec \partial \wedge \vec A(\vec r)=e\sum_{j=1}^N\delta(\vec r-\vec r_j)
\end{equation}
leads to  a magnetic field proportional to the density of particles  in accordance with  (\ref{mag}).
Solving this last equation for $\vec A(\vec r)$ in the Coulomb gauge gives 
\begin{equation}
\vec A(\vec r)= {e\over 2\pi\kappa}\sum_{j=1}^N{\vec k\wedge(\vec r-\vec r_{j})\over (\vec r-\vec r_j)^2}
\end{equation}
in accordance with the A-B potential vector (\ref{A}).
Here  again there is no Lorentz force, the potential vector is a pure gauge, the Chern-Simons term is metric independent, and the field strength is directly related to the matter current.

Coming back to the Hamiltonian formulation (\ref{defbis}), one might ask how the exclusion of the diagonal of the configuration space  materializes in the Hamiltonian  formulation.  One way to look at it is  perturbation theory \cite{Pert1,Pert2}. Let us simplify the problem  by considering the standard A-B problem, i.e. a charged particle in the plane coupled to a flux tube at the origin with the Hamiltonian
\begin{equation}\label{defter}
H={1\over 2m}(\vec p-\alpha{\vec k\wedge\vec r\over r^2})^2 
\end{equation}
Let us see what happens close to Bose statistics when $\alpha\simeq 0$ (by periodicity $\alpha$ can always been chosen in  $[-1/2, +1/2]$, an interval of length $1$,  since in the  one-body case  one quantum of flux can always be gauged away via a regular gauge transformation).
The A-B spectrum \cite{AB}  is   given  by the Bessel functions
 \begin{equation}\label{sol}
\psi(\vec r)=e^{il\theta}J_{|l-\alpha|}(kr)\quad\quad E={k^2\over 2m}
\end{equation}
with wavefunctions vanishing close to  the origin  $r\to 0$  as $J_{|l-\alpha|}(kr)\simeq r^{|l-\alpha|}$. When the angular momentum $l\ne 0$, this is the only possible locally square-integrable function. However, when $l=0$, one could have as well $J_{-|\alpha|}(kr)$ as a  solution, since it is  still locally square-integrable even though it diverges at the origin as $r^{-|\alpha|}$. In principle,  the general solution in the $l=0$ sector should   be a linear combination of $J_{|\alpha|}(kr)$ and $J_{-|\alpha|}(kr)$,  introducing  an additional scale in the coefficient of the linear combination \cite{selfadjoint}.  Restricting the space of solutions as in (\ref{sol}), i.e. wavefunctions vanishing at the origin,  means that a short-range repulsive prescription has been imposed  on the behavior of the wavefunctions when the particle comes close to the flux tube.
One can give a more precise  formulation of this fact by  trying to compute in  perturbation theory the spectrum (\ref{sol}).
Expanding the square in the Hamiltonian  (\ref{defter}), one finds that the $\alpha^2/r^2$ term, which is as singular as the kinetic term,  is divergent at second order in perturbation theory in the $l=0$ sector. It follows that  perturbation theory is not well defined in the problem as defined by the Hamiltonian (\ref{defter}). A renormalization has to be implemented:  one realizes that by adding the countertem $\pi|\alpha|\delta(\vec r)$   to (\ref{defter}), i.e. by considering 
\begin{equation}\label{defquar}
H={1\over 2m}(\vec p-\alpha{\vec k\wedge\vec r\over r^2})^2+ {2\pi|\alpha|\over m}\delta(\vec r)
\end{equation}
the perturbative divergences due to the $\alpha^2/r^2$  term are exactly cancelled by those arising from the 
$\pi|\alpha|\delta(\vec r)$ term at all orders in perturbation theory, giving back the spectrum (\ref{sol}).
Physically, this repulsive $\delta$ contact term means that the particle is prevented from penetrating the core of the flux tube where the field strength is infinite, thus the (at least) $r^{|\alpha|}$ behavior when $r\to 0$.  Note that this has been achieved without introducing any additional scale in the problem.

Clearly, in the $N$-body A-B anyon formulation of the  model,  the corresponding renormalized Hamiltonian should read 
\begin{equation}\label{defquin}
H_N=\sum_{i=1}^N{1\over 2m}(\vec p_i-\alpha\sum_{j\ne i}{\vec k\wedge\vec r_{ij}\over r_{ij}^2})^2+ {2\pi|\alpha|\over m}\sum_{i\ne j}\delta(\vec r_{ij})
\end{equation}
realizing  the quantum mechanical exclusion of the diagonal of the configuration space  in terms of contact repulsive interaction between particles.  Note  that the term $\pi|\alpha|\sum_{i\ne j}\delta(\vec r_{ij})$  in (\ref{defquin}) can also be viewed \cite{Pert2} as  the Pauli spin coupling of the spin of the particles to the singular magnetic field (\ref{mag}) associated  to the flux tubes.

The anyon model defined in (\ref{defquin})  is  properly defined as far as  short-distance  considerations are concerned. It is  
the interacting formulation for regular wavefunctions of the free particles formulation for  multivalued  wavefunctions. Both Hamiltonians  $H_N$ and $H'_N$ are equivalent, the former being more familiar in terms of usual quantum mechanics, the latter more relevant to study braiding and winding properties.

The anyon model  has been the subject of numerous studies in the eighties and the nineties \cite{Myr2}, some of them analytical, starting with the 2-body case which is solvable since its relative 2-body problem is the usual A-B problem (\ref{defter}) with an even (Bose) angular momentum  $l$. The exact solution \cite{LM,W-82-0}  for the relative 2-body problem is given by (\ref{sol}),  $l$ being an even integer, therefore when $\alpha$ is   odd, $l-\alpha$ is odd, corresponding to Fermi statistics (the periodicity $\alpha\to\alpha\pm 2p$ is manifest in the shift $l\to l-\alpha$).
These studies  were followed by the 3-body \cite{Wu1} and then the $N$-body problem \cite{N-body}.  Statistical mechanics was also considered (second virial coefficient \cite{Second,Secondbis}, third virial coefficient \cite{Third}).
However, it soon became  apparent that a complete $N$-body spectrum was out of reach, to the exception of particular classes of exact eigenstates generalizing  the 2-body eigenstates. Numerical \cite{Num} as well as semi-classical \cite{Semi} studies were  performed giving indications on the low energy $N$-body spectrum. A systematic study of the model was  achieved at first \cite{Ouvry1} and  at second order \cite{Ouvry2} in perturbation theory  (at second order the complexity of the model shows up  clearly). Numerical studies \cite{Myr1}, taking some input from the perturbative results, were  performed for the 3rd and 4th virial coefficients. Last but not least, on the experimental side, Laughlin quasiparticles \cite{Laughlin} were put forward as the elementary excitations of  highly-correlated
fractional quantum Hall electron fluids \cite{FQHE}. They were  supposed to carry a fractional charge and  to obey  
anyon statistics \cite{FQHEbis}, a fact confirmed by  Berry phase calculations, at least for quasiholes  \cite{Berry} (for quasiparticles the situation is   less clear). The quasiparticles can propagate quantum-coherently in
chiral edge channels, and constructively or destructively interfere. Unlike
electrons, the interference condition for Laughlin quasiparticles has a
non-vanishing statistical contribution which  might be observed experimentally \cite{Camino}.

Some kind of simplification had to be made to render the model more tractable, and possibly  solvable, at least in a certain sector. One realized that this was the case  if one considered, in addition to  the singular statistical magnetic field, an external homogeneous magnetic field perpendicular to the plane, to which the charge of the  anyons couple.   In the case of a strong magnetic field, by projecting the system of anyons coupled to the magnetic field in its LLL, the model becomes solvable meaning that one can find a class of $N$-body eigenstates which interpolates continuously from the LLL-Bose to the  LLL-Fermi eigenstates basis: this is the LLL-anyon model \cite{Ouvry3}.

\section{The LLL-anyon model}

From now on, let us set the mass of the particles $m=1$ and choose the statistical parameter $\alpha  \in  [-2,0]$. It is understood that all the results below are obtained for $\alpha$ in this interval, but they  can be periodically continued to the whole real  axis.  
Before introducing an  external magnetic field,  let us come back to the anyon Hamiltonian  (\ref{defquin}) and  take advantage of  wavefunctions  vanishing at least as $r_{ij}^{-\alpha}$ when $r_{ij}\to 0$ (exclusion of the diagonal of the configuration space in the quantum mechanical formulation) by encoding this short distance behavior in the $N$-body bosonic wave function \cite{Pert1} 
 \begin{equation}\label{defbisbis}
\psi(\vec r_{1},\vec r_{2},...,\vec r_{N})=\prod_{i<j}r_{ij}^{-\alpha}\tilde{\psi}(\vec r_{1},\vec r_{2},...,\vec r_{N})
\end{equation}
 $\tilde \psi(\vec r_{1},\vec r_{2},...,\vec r_{N})$ is regular but does not have to vanish at coinciding points.
From $H_N$  in (\ref{defquin}) one can compute the new Hamiltonian $\tilde H_N$  acting on $\tilde\psi(\vec r_{1},\vec r_{2},...,\vec r_{N})$.  Since $H_N$ is itself obtained from the free Hamiltonian $H_N'$ in (\ref{free}) via the singular gauge transformation (\ref{def}), it is more transparent to start directly from the free formulation. 
In complex notation (the free Hamiltonian is $H_N'=-2\sum_{i=1}^N\partial\bar\partial_i$) the wavefunction redefinitions (\ref{def}) and (\ref{defbisbis})  combined take  the simple form 
\begin{equation}\label{movebis}
\psi'(z_{1},z_{2},...,z_{N};\bar z_1,\bar z_2,...,\bar z_N)=\prod_{i<j}z_{ij}^{-\alpha}\tilde{\psi}(z_{1},z_{2},...,z_{N};\bar z_1,\bar z_2,...,\bar z_N)
\end{equation} 
The Jastrow-like prefactor $\prod_{i<j}z_{ij}^{-\alpha}$ in (\ref{movebis})  encodes in the wavefunction the essence  of anyon statistics: topological braiding phase  and short-distance contact exclusion behavior. 
It is  immediate that $\tilde H_N$  rewrites as
\begin{equation}\label{final}
\tilde H_N=-2\sum_{i=1}^N\partial_i\bar\partial_i+2\alpha\sum_{i<j}{1\over z_i-z_j}(\bar\partial_i-\bar\partial_j)
\end{equation}
It is a non-Hermitian Hamiltonian (the transformation (\ref{movebis}) is non-unitary), but it has a  simple form,  linear in $\alpha$  and well defined in perturbation theory  (it is perturbatively divergence free). Any analytic  wavefunction of the $z_i$'s  is a $N$-body eigenstate of $\tilde H_N$, and therefore  of the $N$-anyon Hamiltonian (\ref{defquin}) taking into account (\ref{defbisbis}).  Analytical eigenstates  are known to live in the LLL of a magnetic field, if such a field were present. Let us couple the electric charge of each anyon to an external magnetic field $B$  perpendicular to the plane such that by convention $eB>0$ and let us denote  by $\omega_c=eB/2$  half its cyclotron frequency.
One now starts from the Landau Hamiltonian
\begin{equation}\label{B}
H'_N=-{2}\sum_i(\partial_i-{\omega_c\over 2}\bar z_i)(\bar\partial_i+{\omega_c\over 2}z_i)
\end{equation}
In a magnetic field, the $1$-body eigenstates have a long-distance Landau exponential behavior  $\exp(-{1\over 2}\omega_c z_i\bar z_i)$. Let us also encode this behavior in the wavefuction redefinition (\ref{movebis}) so that  it becomes
\begin{equation}\label{moveter}
\psi'(z_{1},z_{2},...,z_{N};\bar z_1,\bar z_2,...,\bar z_N)=\prod_{i<j}z_{ij}^{-\alpha}\exp(-{1\over 2}\omega_c\sum_{i=1}^Nz_i\bar z_i)\tilde{\psi}(z_{1},z_{2},...,z_{N};\bar z_1,\bar z_2,...,\bar z_N)
\end{equation}
One obtains
\begin{equation}\label{finalbis}
\tilde H_N=-2\sum_{i=1}^N(\partial_i\bar\partial_i-\omega_c\bar z_i\bar \partial_i)+2\alpha\sum_{i<j}{1\over z_i-z_j}(\bar\partial_i-\bar\partial_j)+N\omega_c
\end{equation}
where the trivial constant energy shift from the Pauli coupling to the  magnetic field has been ignored. As announced, $\tilde H_N$ acts trivially on $N$-body eigenstates made of symmetrized products of analytic  1-body LLL eigenstates
\begin{equation}\label{LLL}
\sqrt{{\omega_c^{l_i+1}\over \pi l_i!}}z_i^{l_i};\quad l_i\ge 0;\quad E=\omega_c
\end{equation}
(in (\ref{LLL}) the Landau exponential term is missing  since it has already been taken into account in (\ref{moveter})).
So, up to an overall normalization,  
\begin{equation}\label{basis}
\tilde\psi(z_{1},z_{2},...,z_{N};\bar z_1,\bar z_2,...,\bar z_N)={\rm Sym}\,\prod_{i=1}^N z_i^{l_i};\quad  0\le l_1\le l_2\le ...\le l_N 
\end{equation}
is an eigenstate with a degenerate  $N$-body energy, $E_N=N\omega_c$,  a mere reflection of the fact that there are $N$ particles   in the LLL. 
From (\ref{moveter}) and (\ref{basis})  one finally gets 
\begin{equation}\label{theend}
\psi'(z_{1},z_{2},...,z_{N};\bar z_1,\bar z_2,...,\bar z_N)=\prod_{i<j}z_{ij}^{-\alpha}\exp(-{1\over 2}\omega_c\sum_{i=1}^Nz_i\bar z_i){\rm Sym}\prod_{i=1}^N z_i^{l_i};\quad  0\le l_1\le l_2\le ...\le l_N
\end{equation}
The basis (\ref{theend})  continuously interpolates when $\alpha=0\to -1$ from the complete   LLL-Bose $N$-body basis  to the complete LLL-Fermi $N$-body  basis. Indeed, when $\alpha=-1$,  
\begin{equation}\label{}
\psi'(z_{1},z_{2},...,z_{N};\bar z_1,\bar z_2,...,\bar z_N)=\exp(-{1\over 2}\omega_c\sum_{i=1}^Nz_i\bar z_i)\prod_{i<j}z_{ij}{\rm Sym}\,\prod_{i=1}^N z_i^{l_i};\quad  0\le l_1\le l_2\le ...\le l_N 
\end{equation}
 is equivalent to 
\begin{equation}\label{}
\psi'(z_{1},z_{2},...,z_{N};\bar z_1,\bar z_2,...,\bar z_N)=\exp(-{1\over 2}\omega_c\sum_{i=1}^Nz_i\bar z_i){\rm Antisym}\,\prod_{i=1}^N z_i^{l'_i};\quad  0< l'_1< l'_2< ...< l'_N 
\end{equation}
i.e. the LLL fermionic   basis.  One has therefore obtained a complete LLL-Bose $\to$ LLL-Fermi interpolating basis
which allows, in principle, for a complete knowledge of the LLL-anyon system with statistics intermediate between Bose and Fermi statistics. 

One could ask about going beyond the Fermi point $\alpha=-1$ up to the Bose point $\alpha= -2$. This question is related to the validity of the LLL projection, since ignoring higher Landau levels amounts to assuming that excited non LLL states above the $N$-body LLL 
ground state have a non vanishing  gap.  Considerations around the Fermi point, as well as numerical and semiclassical analysis, support \cite{Ouvry3} this scheme as long as $\alpha$ does not come close to $-2$. However, when $\alpha\to -2$, known linear  as well as unknown nonlinear non LLL eigenstates  do join the LLL ground state \cite{Masheigen}. Said differently,  the LLL-anyon basis (\ref{theend})
does not constitute a complete LLL-Bose basis when  $\alpha\to -2$, i.e. some $N$-body LLL bosonic quantum numbers are missing at this point. We will come back to this issue later.

\begin{figure}[htbp]
\epsfxsize=8cm
\centerline{\epsfbox{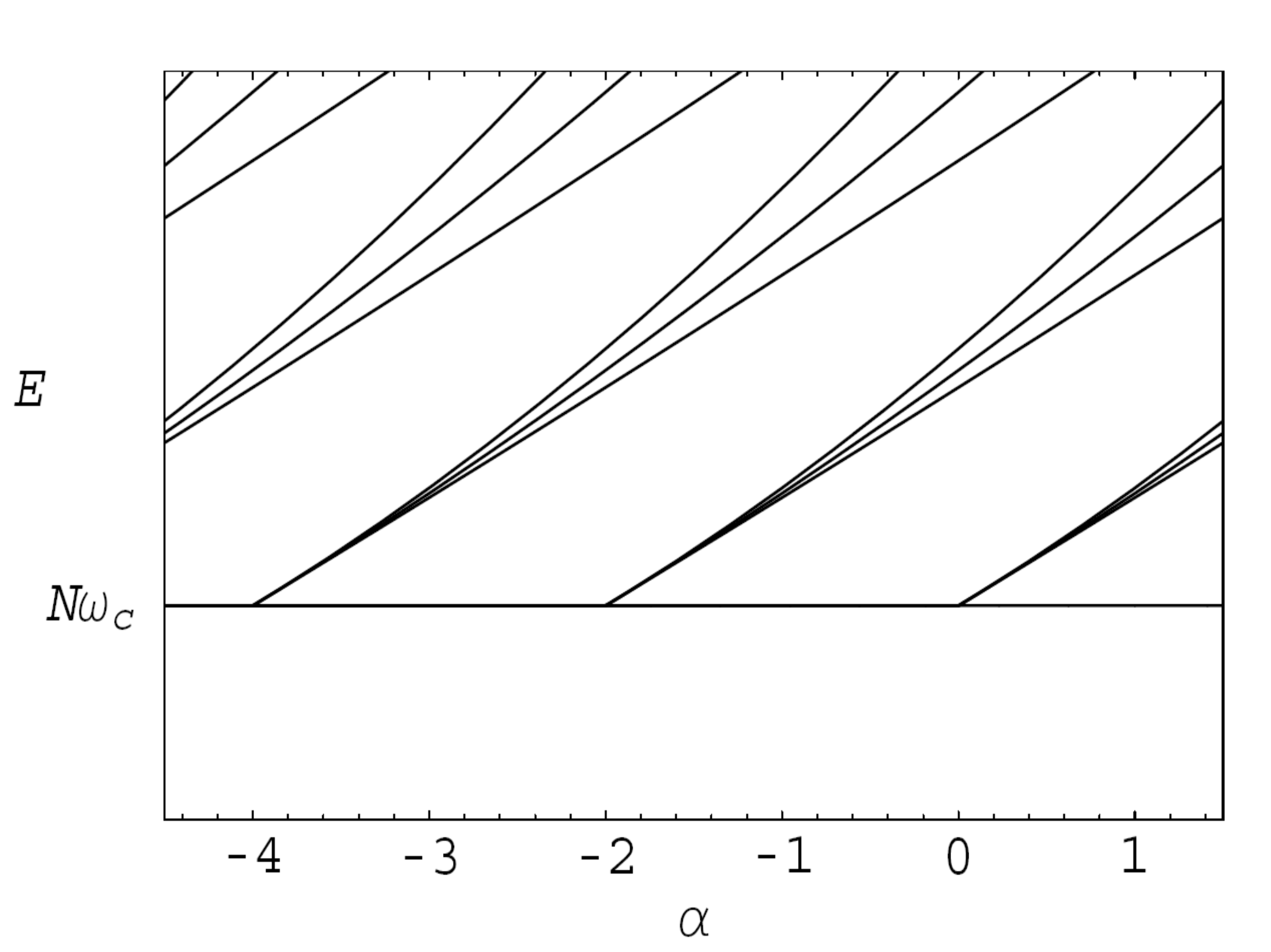}}
\caption{Linear and non linear non LLL eigenstates merge in the LLL ground state at the bosonic values of $\alpha$.}
\label{fig:nonlinear}
\end{figure}

One has not seen yet any $\alpha$ dependence in the $N$-body energy, a situation already encountered in  the 1-body A-B problem,  where the    free continuous energy spectrum  (\ref{sol}) is $\alpha$-independent. This is due to the fact that a magnetic field does not confine particles:  classical orbits are circular cyclotron orbits, but their centers, due to translation invariance, are located anywhere in the plane. Translation invariance in turn gives, in  quantum mechanics, a  Landau spectrum  which is $l_i$ independent, and therefore infinitely degenerate\footnote{From this point of view one can argue that the Landau spectrum is continuous, albeit being made of discrete Landau levels, due   to the   infinite degeneracy on each level.}. The degeneracy factor scales as the infinite surface $V$  of the 2d sample: it is the  flux of the magnetic field counted in units of the flux quantum $\phi_0=2\pi/e$  (in units $\hbar=1$)
\begin{equation}\label{deg}
N_L={VB\over\phi_0}
\end{equation}
Statistical interactions  being topological   interactions, one does not  expect, in the infinite plane limit, any effect  on the $N$-body energies. To see such an effect, one has to introduce a long-distance confinement, like putting the particles in a box. Let us rather introduce \cite{Fermi} a  more convenient  harmonic well confinement where the  particles are trapped, so that the Landau Hamiltonian (\ref{B}) becomes 
 \begin{equation}\label{Bbis}
H'_N=-2\sum_{i=1}^N(\partial_i-{eB\over 4}\bar z_i)(\bar\partial_i+{eB\over 4}z_i)+{1\over 2}\omega^2\sum_{i=1}^Nz_i\bar z_i
\end{equation}
The virtue of the harmonic confinement is to lift the degeneracy  with respect to the angular momentum $l_i$ of the 1-body Landau eigenstates: the harmonic LLL spectrum\footnote{The complete 2d harmonic Landau spectrum is, with the convention $eB>0$,
\begin{equation}\label{red0}
\omega_t(2n_i+l_i+1)-l_i\omega_c;\quad n_i\ge 0,\quad l_i\in Z
\end{equation}
The LLL quantum numbers  are $n_i=0$ and $l_i\ge 0$
.}  becomes  
\begin{equation}\label{LLLomega}
\sqrt{{\omega_t^{l_i+1}\over \pi l_i!}}z_i^{l_i}\exp(-{1\over 2}\omega_t z_i\bar z_i);\quad l_i\ge 0;\quad E=(\omega_t-\omega_c)(l_i+1)+\omega_c
\end{equation}
where $\omega_t=\sqrt{\omega^2+\omega_c^2}$. Each harmonic LLL  level in (\ref{LLLomega}) has now a finite  degeneracy, with an  eigenstate still analytic in $z_i$, up to     the long-distance harmonic Landau combined exponential behavior.
Let us take into account this exponential behavior in the   redefinition of the free $N$-body wavefunction so that  (\ref{moveter}) now becomes
\begin{equation}\label{movequar}
\psi'(z_{1},z_{2},...,z_{N};\bar z_1,\bar z_2,...,\bar z_N)=\prod_{i<j}z_{ij}^{-\alpha}\exp(-{1\over 2}\omega_t\sum_{i=1}^Nz_i\bar z_i)\tilde{\psi}(z_{1},z_{2},...,z_{N};\bar z_1,\bar z_2,...,\bar z_N)
\end{equation}
Starting from the  Hamiltonian (\ref{Bbis}) one obtains \cite{Ouvry4, Ouvry4bis} 
\begin{equation}\label{finalter}
\tilde H_N=-2\sum_{i=1}^N(\partial_i\bar\partial_i-{\omega_t+\omega_c\over 2}\bar z_i\bar \partial_i -{\omega_t-\omega_c\over 2} z_i \partial_i)+2\alpha\sum_{i<j}\big[{1\over z_i-z_j}(\bar\partial_i-\bar\partial_j)-{\omega_t-\omega_c\over 2}\big]  +N\omega_c
\end{equation}
Again let us act on $N$-body eigenstates made,  in analogy with (\ref{basis}), of symmetrized products of  the 1-body harmonic LLL  eigenstates (\ref{LLLomega})
\begin{equation}\label{}
\tilde\psi(z_{1},z_{2},...,z_{N};\bar z_1,\bar z_2,...,\bar z_N)={\rm Sym}\,\prod_{i=1}^N z_i^{l_i};\quad  0\le l_1\le l_2\le ...\le l_N 
\end{equation}
Acting on this basis, the Hamiltonian (\ref{finalter}) rewrites as
 \begin{equation}\label{}
{\tilde H_N}=({\omega_t-\omega_c})\big[\sum_{i=1}^N z_i \partial_i-\alpha{N(N-1)\over 2}+N\big]  +N\omega_c
\end{equation}
so that the  $N$-anyon energy  spectrum is
\begin{equation}\label{energy}
E_N= ({\omega_t-\omega_c})\big[\sum_{i=1}^N l_i -\alpha{N(N-1)\over 2}+N\big]  +N\omega_c
\end{equation}
The $N$-anyon spectrum (\ref{energy}) is  a sum of  1-body harmonic LLL spectra  shifted by the 2-body statistical term $-({\omega_t-\omega_c})\alpha{N(N-1)/ 2}$. The effect of the harmonic well  has been not only  to lift the degeneracy with respect to the $l_i$'s, but also to make  the energy dependence on $\alpha$ explicit.
When computing 
thermodynamical quantities like the equation of state, the harmonic well regulator  will also  be needed
to compute finite quantities in a finite ``harmonic'' box, and then take the thermodynamic limit, by letting $\omega\to 0$ in an appropriate way.

The resulting eigenstates from (\ref{movequar}) 
\begin{equation}\label{etoui}
\psi'(z_{1},z_{2},...,z_{N};\bar z_1,\bar z_2,...,\bar z_N)=\prod_{i<j}z_{ij}^{-\alpha}\exp(-{1\over 2}\omega_t\sum_{i=1}^Nz_i\bar z_i){\rm Sym}\,\prod_{i=1}^N z_i^{l_i};\quad  0\le l_1\le l_2\le ...\le l_N 
\end{equation}
are called ''linear states''  since their energy (\ref{energy}) varies linearly with $\alpha$. 
As already  stressed, they constitute a set of exact $N$-body eigenstates  which is  only a  small part of the complete $N$-body spectrum, which  remains mostly unknown. However, what makes, in  the LLL  context, these linear  states particularly interesting  is  that they continuously interpolate when $\alpha=0\to-1$ from the complete harmonic LLL-Bose  basis to the complete harmonic LLL-Fermi   basis.

Before turning to LLL-anyon thermodynamics,
let us reconsider  the physical charge-flux composite interpretation of the anyon model, where the charges are now coupled to  an external  magnetic field. A given particle, say the $N$th, sees  a ``positive'' ($eB>0$)  magnetic field perpendicular to the plane, and $N-1$ ``negative'' ($e\phi=2\pi\alpha<0$, $\alpha\in [-1,0]$) point vortices piercing the plane at the positions of the other particles.  This is a screening regime: in the large $N$ limit where a mean field picture is expected to be valid, the more $\alpha$ is close to the fermionic point $\alpha=-1$, the more the external magnetic field is screened by the mean magnetic field associated with the vortices. In terms of the  total (external + mean) magnetic  field $\langle B\rangle$  that the $N$th particle sees, or rather in terms of its  flux $V\langle B\rangle$, or, when counted in units of the flux quantum, in terms of the Landau degeneracy $\langle N_L\rangle$, one has 
 \begin{equation}\label{haldane}
{V\langle B\rangle/\phi_0}= {(VB)/\phi_0} + {(N-1)\phi/ \phi_0} \quad{\rm i.e.}\quad 
\langle N_L\rangle=N_L+(N-1)\alpha\end{equation}
Moving away from the Bose point, i.e. $\alpha\le 0$,  as $N$ increases  the number $\langle N_L\rangle$ of 1-body quantum states available for the $N$th particle in the LLL of $\langle B \rangle$ decreases. This sounds reasonable, bearing in mind that a fermion occupies a  quantum state  to the exclusion of others  (Pauli exclusion), whereas bosons can condense (Bose condensation). Introducing the LLL filling factor
\begin{equation}\label{nu}
\nu= {N\over N_L}
\end{equation}
one  deduces from (\ref{haldane}) a maximal critical filling \cite{Ouvry3}  for which the screening is total, $\langle N_L\rangle=0$
\begin{equation}\label{nubis}
\nu= -{1\over \alpha}
\end{equation}
This is nothing but recognizing once more  that  bosons ($\alpha=0$) can infinitely fill a quantum state ($\nu=\infty$), whereas  fermions ($\alpha=-1$) are at most one per quantum state ($\nu=1$). In between, one finds that there are at most $-1/\alpha$ anyons per quantum state.
 
Interestingly enough,    Haldane/exclusion statistics definition\footnote{This is Haldane's statistics  for one particle species. It can be generalized to the multispecies case.} happens to coincide with (\ref{haldane}):  for a gas of particles obeying  Haldane/exclusion statistics \cite{Haldane} with statistical parameter $g\in [0,1]$,  given  $N_L$ degenerate energy levels  and $N-1$ particles already populating the levels, the number $d_N$ of quantum states  still available for the $N$th particle is given by (\ref{haldane})  where $-\alpha$ is replaced by $g$
\begin{equation}\label{haldanebis}d_N=N_L-(N-1)g
\end{equation}
On the one hand, Haldane's definition (\ref{haldanebis}) stems from an arbitrary  combinatorial point of view, inspired by the Bose and Fermi counting of states. On the other hand,  in the LLL-anyon model, (\ref{haldane})  is obtained from a somehow ad-hoc  mean field ansatz.  We will come back to these issues in the next section.

\section{LLL-anyon thermodynamics}

Let us rewrite the $N$-body energy (\ref{energy}) as \cite{Ouvry5}
\begin{equation}\label{energygen}
E_N= \sum_{i=1}^N(\epsilon_0 + l_i\tilde\omega) -\alpha{N(N-1)\over 2}\tilde\omega;\quad 0\le l_1\le l_2\le...\le l_N
\end{equation}
with $\tilde\omega=({\omega_t-\omega_c})$ and $\epsilon_0=\omega_c$. Introducing the fugacity $z$  and the inverse temperature $\beta$, one wants to compute the thermodynamic potential   
\begin{equation}\label{thermo}
\ln Z(\beta, z)= \ln (\sum_{N=0}^{\infty} z^NZ_N);\quad Z_0=1
\end{equation}
where $Z(\beta,z)$ is the  grand partition function defined in terms of the $N$-body partition functions
$Z_N=\mathrm{Tr}\exp(-\beta H'_N)=\mathrm{Tr}\exp(-\beta H_N)=\mathrm{Tr}\exp(-\beta {\tilde H}_N)$.
The thermodynamic potential rewrites  as $\ln Z(\beta, z)=\sum_{n=1}^{\infty} b_n z^n$ where, at order $z^n$, the cluster coefficient $b_n$ only requires the knowledge of the  $Z_i$'s, with $i\le n$. One is interested in evaluating the thermodynamic potential in the thermodynamical limit, i.e. $\omega$ is small, which means, here, that the dimensionless quantity $\beta\omega$ is small. The $N$-body spectrum, as given in (\ref{energygen}), allows  to  compute, at leading order in $\beta\omega\to 0$, the $Z_i$'s for $i\le n$, and thus the $ b_n$'s
\begin{equation}\label{thermobis}
b_n={1\over \beta\tilde\omega}{e^{-n\beta\omega_c}\over n^2}\prod_{k=1}^{n-1}{k+n\alpha\over k};\quad b_1={1\over \beta\tilde\omega}e^{-\beta\omega_c}
\end{equation}
One has still to give a meaning, in the thermodynamic limit $\beta\omega= 0$,   to the scaling factor $1/(\beta\tilde\omega)$
in (\ref{thermobis}). To this purpose, one temporarily switches off the anyonic interaction and the external magnetic field, and considers a quantum gas  of  non interacting harmonic oscillators per se. One asks, when $\beta\omega\to 0$, for its cluster coefficients to yield the infinite box (plane wave) cluster coefficients. At order $n$ in the cluster expansion, in $d$ dimensions, one obtains \cite{Pert1}
\begin{equation}\label{thermolimit}
\lim_{\beta\omega\to 0}({1\over n(\beta\omega)^2})^{d\over 2}= {V\over \lambda ^d}
\end{equation}
where  $\lambda=\sqrt{2\pi\beta}$ is the thermal wavelength and $V$ is  the  $d$-dimensional infinite volume (in $d=2$ dimensions,  $V$ is, as  defined above,  the infinite area of the 2d sample).
Using the thermodynamic limit prescription (\ref{thermolimit}),  the cluster coefficient (\ref{thermobis}) rewrites, in the thermodynamic limit, as \cite{Ouvry3}
 \begin{equation}\label{thermoter}
b_n=N_L{e^{-n\beta\omega_c}\over n}\prod_{k=1}^{n-1}{k+n\alpha\over k};\quad b_1=N_Le^{-\beta\omega_c}
\end{equation}
The cluster expansion $\ln Z(\beta,z)=\sum_{n=1}^{\infty} b_n z^n$, as a power series of  $ze^{-\beta\omega_c}<1$, can be summed up
\begin{equation}\label{thermoquar}
\ln Z(\beta,z)=N_L\ln y(ze^{-\beta\omega_c})
\end{equation}
 where $y(ze^{-\beta\omega_c})$, a function of the variable $ze^{-\beta\omega_c}$, is such that
\begin{equation}\label{thermoquarbis}
\ln y=ze^{-\beta\omega_c}+\sum_{n=2}^{\infty}{(ze^{-\beta\omega_c})^n\over n}\prod_{k=1}^{n-1}{k+n\alpha\over k}
\end{equation}
It obeys  \cite{Ouvry3} 
\begin{equation}\label{thermoquin}
y- ze^{-\beta\omega_c}y^{1+\alpha}=1
\end{equation}
and  has in turn a power series expansion \cite{Poly} 
 \begin{equation}\label{thermoquinbis}
y=1+ze^{-\beta\omega_c}+\sum_{n=2}^{\infty}{(ze^{-\beta\omega_c})^n}\prod_{k=2}^{n}{k+n\alpha\over k}
\end{equation}
 From (\ref{thermoquar}) one  infers that $Z(\beta,z)=y^{N_L}$  so that \cite{Ouvry4,Bytsko}
\begin{equation}\label{thermosix}
Z(\beta,z)=y^{N_L}=1+N_Lze^{-\beta\omega_c}+N_L\sum_{N=2}^{\infty}{(ze^{-\beta\omega_c})^N}\prod_{k=2}^{N}{k+N_L+N\alpha-1\over k}
\end{equation}
Clearly, from (\ref{thermosix}), the $N$-body partition function $Z_N$ is
\begin{equation}\label{thermoseven}
Z_N=N_L{e^{-N\beta\omega_c}}\prod_{k=2}^{N}{k+N_L+N\alpha-1\over k}
\end{equation}
It is, by construction, positive. Necessarily, $\alpha$ and $N_L$ being given, $N$ has to be such that $N_L+N\alpha\ge 0$. This  always is the case as long as $N$ is finite, since $N_L$ scales like the infinite surface of the 2d sample. In the thermodynamic limit, where $N\to \infty$, the condition $N_L+N\alpha\ge 0$ implies 
for the filling factor
\begin{equation}\label{mirac}
\nu\le -{1\over \alpha}
\end{equation}
It is rather striking that the RHS of (\ref{mirac}), which has just been derived from the exact computation of the cluster coefficients from  the $N$-body spectrum, is nothing but the critical filling (\ref{nubis}) obtained in the mean field approach when the screening is total.

The ``degeneracy`` associated with  $N$ anyons populating   the LLL quantum states is, from (\ref{thermoseven}),
\begin{equation}\label{degeneracy}
{N_L}\prod_{k=2}^{N}{k+N_L+N\alpha-1\over k}={N_L\over N!}{(N+N_L+N\alpha-1)!\over (N_L+N\alpha)!}
\end{equation}
where  a factorial  with a negative argument has to be understood  as $(-p)!=\lim_{x\to 0} (-p+x)!$.

When $\alpha=0$, this is the usual Bose counting factor for the  number of ways to put $N$ bosons  in $N_L$ states  
\begin{equation}\label{degeneracybose}
{(N+N_L-1)!\over N!(N_L-1)!}
\end{equation}
When $\alpha=-1$, this is the Fermi counting  factor ${N_L!/(N!(N_L-N)!)}$. 
If one   considers for a moment the statistical parameter to be a negative integer $\alpha\le -1 $ , 
the degeneracy (\ref{degeneracy}) still allows for a  combinatorial interpretation \cite{Poly} : provided again that $N_L+N\alpha\ge 0$, it is   the number of ways to put  $N$ particles on  a  circle consisting of  $N_L$  quantum states  such that there are at least $-\alpha-1$ empty  states in between two occupied states. When $\alpha=-1$, this is nothing but the usual exclusion mechanism  for fermions (one fermion at most per quantum state). When $\alpha\le -1$, i.e. beyond the Fermi point, more and more states are excluded between  two filled states. In the case of interest $\alpha$ in $[-1,0]$, one has a ''fractional`` exclusion where one can put more than one particle per quantum state according to the fractional $\alpha$, but not  infinitely many as in the Bose case.

The degeneracy (\ref{degeneracy}) originates  from  the exact $N$-body spectrum (\ref{energy}).  In the case of Haldane statistics as defined in (\ref{haldanebis}), there is no Hamiltonian and no $N$-body spectrum to begin with. One  rather starts from the Bose counting factor (\ref{degeneracybose}) 
and bluntly replaces, in accordance with  (\ref{haldanebis}), $N_L$  by $N_L-(N-1)g$  to obtain 
\begin{equation}\label{degeneracybis}
{(N_L-(N-1)(g-1))!\over N!(N_L-(N-1)g-1)!};
\end{equation}
which  indeed  interpolates,  when $g=1$, to the Fermi counting factor. The degeneracy (\ref{degeneracybis}) is similar to (\ref{degeneracy}): if one allows the exclusion parameter $g$ to be  an integer, it counts \cite{Poly} the number of ways to put $N$ particles on a line of finite length consisting of $N_L$ quantum states such that there are at least $g-1$ empty  states in between two occupied states. Up to boundary conditions on the space of available quantum states (periodic  versus infinite  wall), both counting (\ref{degeneracy}, \ref{degeneracybis}) are identical. In the thermodynamic limit when $N$ becomes large, boundary conditions should not play a role anymore: not surprisingly,
starting from (\ref{degeneracybis}) and following the usual  route of statistical mechanics \cite{Wu2} (saddle-point approximation) leads, in the thermodynamic limit, to the same LLL-anyon thermodynamic potential given by the equations  (\ref{thermoquar}) and (\ref{thermoquin}),  where the anyonic parameter $-\alpha$ is replaced by  the exclusion parameter $g$.

Note that the grand partition factorization  $Z(\beta,z)=y^{N_L}$ in (\ref{thermoquar})   could suggest \cite{Wilczek2} an interpretation of  $y$ as a  LLL-anyon  grand-partition function  for a single quantum state at energy $\omega_c$,  on the same footing as, when $\alpha=0$ or $\alpha=-1$,  $y=(1\mp ze^{-\beta\omega_c})^{\mp 1}$  is indeed the single quantum state grand partition function  for a Bose or Fermi gas. This interpretation is  not possible for the reason advocated above: it would yield, as soon as $\alpha$  is fractional, negative $N$-body partition functions. This is clearly impossible: the $N$-body anyonic system is, except in the Bose and Fermi cases, truly interacting and therefore its statistical mechanics is by no means factorisable to a single-state  statistical mechanics.

From (\ref{thermoquar}, \ref{thermoquin}), the average energy  $\bar E\equiv -{\partial \ln Z(\beta,z)/ \partial \beta}$ and the average particle number $\bar N \equiv z {\partial \ln Z(\beta,z)/ \partial z}$ or, equivalently,  
the filling factor  $\nu=\bar N/N_L$,  can be computed. $\nu$ satisfies 
\begin{equation} \label{central}
y=1+ {\nu\over 1+\alpha\nu}
\end{equation}
or, equivalently, using (\ref{thermoquin})
\begin{equation} \label{centralbis}
ze^{-\beta\omega_c}={\nu\over (1+(1+\alpha)\nu)^{1+\alpha}(1+\alpha\nu)^{-\alpha}}
\end{equation}
When $\alpha\ne 0$ and $\alpha\ne-1$, this equation cannot in general be solved analytically, except in special cases like  $\alpha=-1/2$ (semions). The equation of state follows
\begin{equation} \label{centralter}
\beta P V=\ln(1+ {\nu\over 1+\alpha\nu})
\end{equation}
In all these equations, it is  understood from (\ref{mirac}) that  $\nu\le -{1/ \alpha}$. When $\nu=-{1/ \alpha}$, the pressure diverges, a  manifestation of the fact that there are as many anyons as possible in the LLL,  higher Landau levels being forbidden by construction.
One also notes that, for the degenerate LLL gas, the filling factor in (\ref{centralbis}) is nothing but the mean occupation number $n$ at  energy $\epsilon=\omega_c$ and fugacity $z$. As
expected, (\ref{centralbis})  at $\alpha=0$  gives the standard Bose  mean occupation number $n=ze^{-\beta\epsilon}/(1-ze^{-\beta\epsilon})$, whereas at $\alpha=-1$ it gives  the Fermi mean occupation number $n=ze^{-\beta\epsilon}/(1+ze^{-\beta\epsilon})$.

The entropy $S\equiv \ln Z(\beta,z)+\beta\bar E-(\ln z) \bar N$  is (trivially $\bar E=\bar N\omega_c$ since the $N$ particles are in the LLL)
\begin{equation} \label{centralquar}
S=N_L\big[(1+\nu(1+\alpha))\ln(1+\nu(1+\alpha))-(1+\nu\alpha)\ln(1+\nu\alpha)-\nu\ln\nu\big]
\end{equation}
It  vanishes when  $\nu=-{1/ \alpha}$, an indication that the $N$-body LLL anyon  eigenstate is not degenerate at the critical filling. From (\ref{energy}), one infers that the $N$-body eigenstate of lowest energy has all its one-body orbital momenta quantum numbers $l_i=0$. It follows from (\ref{theend}) that, in the thermodynamic limit  at the critical filling, the LLL-anyon non-degenerate groundstate wavefunction is 
\begin{equation}\label{theendbis}
\psi'(z_{1},z_{2},...,z_{N};\bar z_1,\bar z_2,...,\bar z_N)=\prod_{i<j}z_{ij}^{-\alpha}\exp(-{1\over 2}\omega_c\sum_{i=1}^Nz_i\bar z_i);\quad \nu=-{1\over \alpha}
\end{equation}
with total angular momentum 
 \begin{equation}\label{}
L={N(N-1)\over 2\nu}
\end{equation}
The pattern in (\ref{theendbis}) is reminiscent of the Laughlin wavefunctions at  FQHE fillings  $\nu=1/(2m+1)$
 \begin{equation}\label{Laughlin}
\psi(z_{1},z_{2},...,z_{N};\bar z_1,\bar z_2,...,\bar z_N)=\prod_{i<j}z_{ij}^{2m+1}\exp(-{1\over 2}\omega_c\sum_{i=1}^Nz_i\bar z_i);\quad \nu={1\over 2m+1}
\end{equation}
On the one hand, Laughlin wavefunctions are fermionic, their filling factors are rational numbers smaller than 1, and they are approximate solutions to the  underlying $N$-body  Coulomb dynamics in a strong magnetic field. On the other hand,  LLL-anyon wavefunctions are multivalued, their filling factor continuously interpolates between $\infty$ and $1$, and they are exact solutions to the $N$-body LLL anyon problem.  Still, the similarity between (\ref{theendbis}) and (\ref{Laughlin}) is striking.

Trying to push (\ref{theendbis}) further beyond the Fermi point  eventually   up to the Bose point at $\alpha=-2$, one obtains  a Bose gas at filling $\nu=1/2$ with the non-degenerate wavefunction
\begin{equation}\label{theendter}
\psi'(z_{1},z_{2},...,z_{N};\bar z_1,\bar z_2,...,\bar z_N)=\prod_{i<j}z_{ij}^{2}\exp(-{1\over 2}\omega_c\sum_{i=1}^Nz_i\bar z_i);\quad \nu={1\over 2}
\end{equation}
One already knows that the LLL-anyon basis (\ref{theend})  is not interpolating to the complete LLL-Bose basis when $\alpha= -2$. At this point,  non LLL $N$-body eigenstates merge in the LLL ground state to compensate  for some missing bosonic quantum numbers -see Figure \ref{fig:nonlinear}.  Clearly, (\ref{theendter}) should  reproduce, by periodicity,  the bosonic non-degenerate wavefunction  (\ref{theendbis}) at $\alpha=0$, but it does not. On the same footing, when $\alpha=-2$ the critical filling should be bosonic, i.e. $\nu=\infty$, whereas  $\nu=1/2$. The  unphysical critical filling discontinuity, $\infty$ versus $1/2$,  is yet another manifestation of the missing bosonic quantum numbers. In other words, the very eigenstates which join the LLL ground state at the Bose point $\alpha=-2$  and   provide for the missing  quantum numbers, have the effect to smooth out the critical filling discontinuity. Still, it has been  shown \cite{Anti} that the stronger the magnetic field $B$ is, the more valid (\ref{theendbis})  remains closer and closer to $\alpha=-2$. The limit  $\alpha\to -2$  is,  due to periodicity, the same as the limit $\alpha\to 0$  from above, which can be described as  an anti-screening regime.  One concludes  that close to the Bose point $\alpha=0$, the critical filling of a LLL-anyon  gas is $\nu=\infty$  or $\nu=1/2$ depending    on infinitesimally moving away from the Bose point in the screening regime (the ground state wavefunction is the usual non degenerate bosonic wavefunction),  or in the anti-screening regime (the ground state wavefunction is  (\ref{theendter})). Again, the Bose point has a somehow singular behavior, a feature already encountered in  perturbation theory. Note finally that the occurrence of the  $\nu=1/2$ fraction for the bosonic filling factor in the antiscreening regime is  physically challenging: fast rotating Bose-Einstein condensates in the FQHE regime  are expected \cite{Bose} to reach a $1/2$ filling described by  the Laughlin-like  wavefunction (\ref{theendter}).

\begin{figure}[htbp]
\epsfxsize=8cm
\centerline{\epsfbox{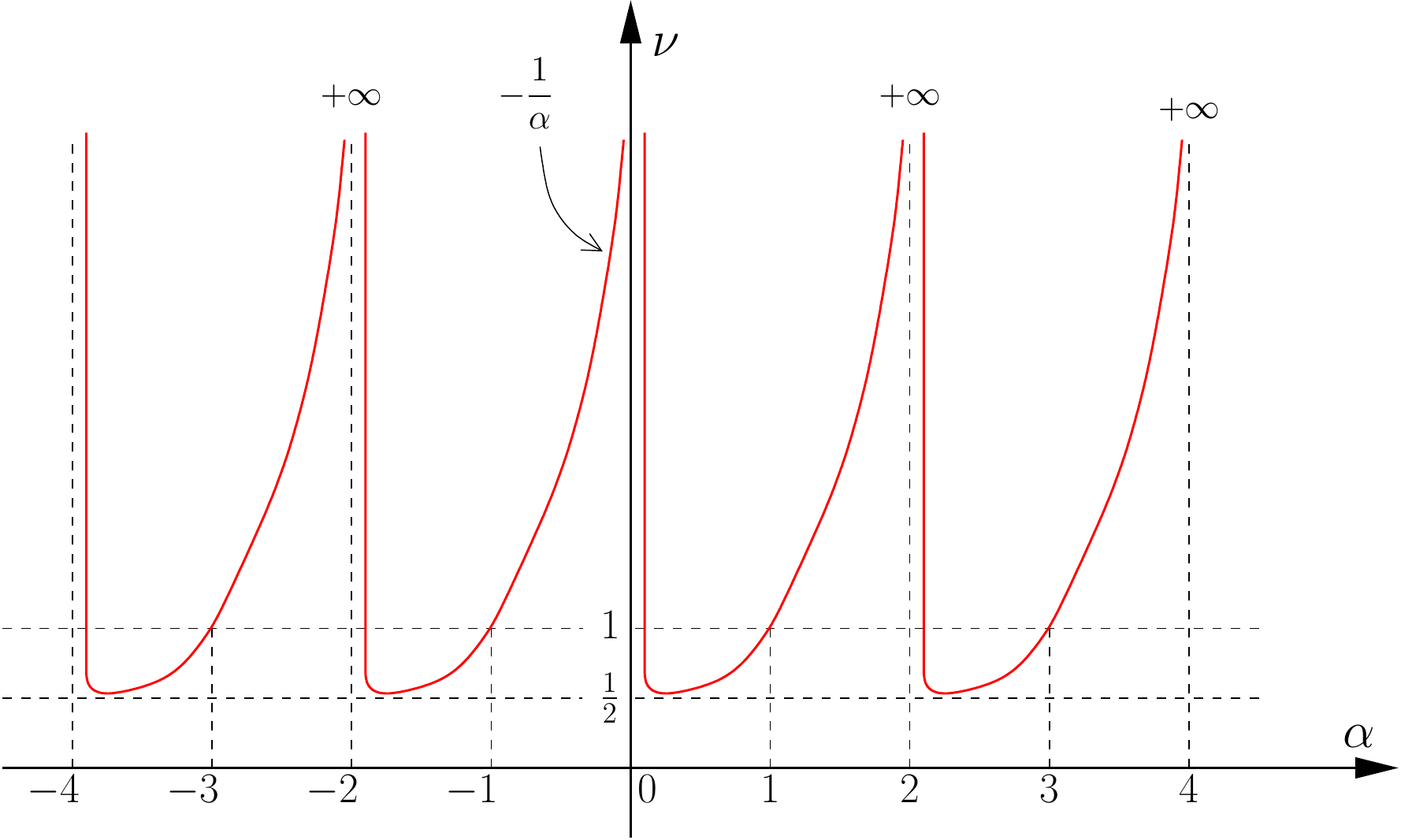}}
\caption{The critical LLL-anyon filling curve as a function of $\alpha$. The critical Bose filling $\nu={1\over 2}$ occurs at the Bose points in the anti-screening regime. The continuity of the critical curve at these points is restaured by the non LLL eigenstates joining the LLL ground state.}
\label{fig:anti}
\end{figure}

So far one has been concerned with  two-dimensional  systems: in the thermodynamic limit, a single particle in the LLL, and, consequently,  a gas of LLL-anyons, are two dimensional, as can be seen   from the $N_L\simeq V$  scalings\footnote{In the LLL there is only one quantum number $l_i$ per particle, still the system is 2d.} of the 1-body LLL partition function $Z_{LLL}=N_L\exp(-\beta\omega_c)$  and  the LLL anyon thermodynamic    potential (\ref{thermoquar}). Denoting by  $\rho_{LLL}(\epsilon)=N_L\delta(\epsilon-\omega_c)$  the 1-body LLL density of states, (\ref{thermoquar}) can be rewritten as 
\begin{equation}\label{thermoquarter}
\ln Z(\beta,z)=\int_0^{\infty}\rho_{LLL}(\epsilon)\ln y(ze^{-\beta\epsilon})d\epsilon
\end{equation}
Convincingly, in  (\ref{thermoquarter}) the one-body dynamics of individual particles is described by the one-body density of states, whereas the LLL anyon statistical collective behavior is encoded in  the $y$  function which depends on the statistical parameter $\alpha$.  

One might ask  about  other integrable $N$-body systems which would lead to the same kind  of statistics. It would be  tempting to define a model obeying  fractional/exclusion statistics if,  its  one-body density of states  $\rho(\epsilon)$  being given, its thermodynamic potential  has  the form 
\begin{equation}\label{thermoquarquar}
\ln Z(\beta,z)=\int_0^{\infty}\rho(\epsilon)\ln y(ze^{-\beta\epsilon})d\epsilon
\end{equation} with
\begin{equation}\label{thermoquarquin}
y- ze^{-\beta\epsilon}y^{1+\alpha}=1
 \end{equation}
so that
\begin{equation}\label{thermoquinbisbisbis}
y=1+ze^{-\beta\epsilon}+\sum_{n=2}^{\infty}{(ze^{-\beta\epsilon})^n}\prod_{k=2}^{n}{k+n\alpha\over k}
\end{equation}
The
  mean occupation number follows as $n=z\partial \ln y/\partial z$. It obeys
\begin{equation} \label{gen}
y=1+ {n\over 1+\alpha n};\quad{\rm or}\quad   n={y-1\over 1-\alpha(y-1)}\ge 0
\end{equation}
or, equivalently,
\begin{equation} \label{genbis}
ze^{-\beta\epsilon}={n\over (1+(1+\alpha)n)^{1+\alpha}(1+\alpha n)^{-\alpha}}
\end{equation}
One has the duality relation \cite{Wilczek2}
\begin{equation} \label{duality}
1={1\over y} +{1\over \tilde {y}}; \quad{\rm where}\quad  \tilde{y}- ({z}e^{-\beta\epsilon})^{-1}{\tilde {y}}^{1+{1\over\alpha}}=1
\end{equation}
or, equivalently
\begin{equation} \label{dualitybis}
-\alpha n -{1\over \alpha}\tilde{n}=1
\end{equation}
where $\tilde n$ is related to $\tilde y$ as $n$ to $y$ in (\ref{gen}). The duality relation (\ref{duality},\ref{dualitybis}) can be interpreted as  a particle-hole symmetry  relation. 
Setting $t=ze^{-\beta\omega_c}$, one also has a simple expression 
\cite{Italiens} 
for $dn(t)/ dt$
\begin{equation} \label{centralbisbis}
t{dn\over dt}=n(1+(1+\alpha)n )(1+\alpha n)
\end{equation}

All these equations have been  understood  as arising microscopically from the LLL anyon Hamiltonian  with 
one-body density of states $\rho(\epsilon)=\rho_{LLL}(\epsilon)$. It happens that it is possible to find another $N$-body microscopic Hamiltonian which  leads to the thermodynamics  (\ref{thermoquarquar}).
Consider, in one dimension, the integrable $N$-body Calogero model \cite{Calogero}  with inverse-square 2-body interactions
\begin{equation}
{ H_N}= -\frac{1}{2}\sum_{i=}^N \frac{\partial^2}{\partial x_i^2} + 
\alpha(1+\alpha)\sum_{i<j}\frac{1}{(x_i-x_j)^2}
+\frac{1}{2}\,\omega^2\, \sum_{i=1}^N x_i^2
\label{calogero}
\end{equation}
where $x_i$ represents the position of the $i$-th particle on the infinite 1d line.
This model is known to describe particles with nontrivial statistics in one dimension interpolating from Bose ($\alpha=0$) to Fermi ($\alpha=-1$) statistics. It means that the $1/x^2$ Calogero interaction is purely statistical, without any classical effect on  particle  motions, up to a overall reshuffling of the particles \cite{Calogero1}. The Calogero model  remains integrable when, as in (\ref{calogero}), a confining 1d harmonic well is added. This is the harmonic Calogero model, whereas the Calogero-Sutherland model \cite{Sutherland} would have the particles confined on  a circle. The effect of the harmonic well is, as in the LLL anyon case,  to lift the thermodynamic limit degeneracy  in such a way that  the $N$-body harmonic Calogero spectrum ends up depending on the Calogero coupling constant $\alpha$
\begin{equation}\label{energycal}
E_N= \omega\big[\sum_{i=1}^N l_i -\alpha{N(N-1)\over 2}+{N\over 2}\big];\quad 0\le l_1\le l_2\le ...\le l_N
\end{equation}
Here the $l_i$'s  correspond to  the quantum numbers of the  1-d harmonic Hermite polynomials free 1-body eigenstates
\begin{equation}\label{harspec}
({\omega\over \pi})^{1/4}{1\over \sqrt{2^{l_i} l_i!}}e^{-{1\over2}\omega x_i^2}H_{l_i}(\sqrt{\omega}x_i);\quad l_i\ge 0;
  \quad E= \omega(l_i +{1\over 2})
\end{equation}

It is remarkable that (\ref{energycal}) happens to be again of the form  (\ref{energygen})  with $\tilde{\omega}=\omega$, $\epsilon_0=\omega/2$.   
Following the same steps as in the LLL-anyon case, and using again (\ref{thermolimit}) while taking the thermodynamic limit $\beta\omega\to 0$, the Calogero cluster coefficients rewrite as
 \begin{equation}\label{thermocal}
b_n={L\over \lambda}{1\over n\sqrt n}\prod_{k=1}^{n-1}{k+n\alpha\over k};\quad b_1={L\over \lambda}
\end{equation}
where the infinite length of the 1d line  has been denoted by $L$.     The cluster expansion 
can still be resumed using (\ref{thermoquarbis}) provided the unwanted $1/\sqrt n$ term in (\ref{thermocal}) is properly taken care of. Introducing the 1d plane wave momentum $k$ 
 \begin{equation}\label{}
{1\over \lambda\sqrt{n}}={1\over 2\pi}\int_{-\infty}^{\infty} dk  e^{-n\beta {k^2\over 2}} 
\end{equation}
and denoting the 1-body energy as $\epsilon= {k^2/ 2}$, one finally obtains 
\begin{equation}\label{thermocalbis}
\ln Z(\beta,z)=\int_0^{\infty}\rho_0(\epsilon)\ln y(ze^{-\beta\epsilon})d\epsilon
\end{equation}
where 
\begin{equation}
\rho_0(\epsilon)={L\over\pi\sqrt{2\epsilon}}
\end{equation}
is the free 1-body density of states in one dimension. This is not a surprise: in the thermodynamic limit $\omega\to 0$, where $l_i\to\infty$ with $l_i\omega=k_i^2/2$  kept fixed,  the Hermite polynomial $H_{l_i}$  becomes a plane wave of momentum $k_i$.

From (\ref{thermocalbis}), one concludes\footnote{The same conclusion  would be reached starting form the Calogero-Sutherland  model and taking the corresponding thermodynamic limit, i.e. the radius of the confining circle going to infinity.}
that, in the thermodynamic limit, the Calogero model has indeed a LLL-anyon/exclusion like statistics  \cite{Isakov} according to (\ref{thermoquarquar}) and (\ref{thermoquarquin}), interpolating, as it should, from  a free bosonic 1d gas at $\alpha=0$ to   a free fermionic 1d gas at $\alpha=-1$.

It follows that the $2$d LLL-anyon and $1$d Calogero models,  which seem a priori  unrelated, do obey  the same type of statistics. This is not a coincidence. Looking at their harmonic $N$-body spectrum  (\ref{energy}) and (\ref{energycal}), one realizes that, up to an irrelevant zero-point energy,  the latter is  the $B\to 0$ limit of the former. This  remains true in the thermodynamic limit $\omega\to 0$. So, not only (\ref{thermoquarter}) and 
(\ref{thermocalbis}) are of the same type, but also, when $B\to0$, (\ref{thermoquarter})  has to become
(\ref{thermocalbis}). It follows that,  necessarily, the 1-body densities of states $\rho_{LLL}(\epsilon)$  and $ \rho_{0}(\epsilon)$ satisfy  $\lim_{B\to 0} \rho_{LLL}(\epsilon)=\rho_{0}(\epsilon)$, i.e.
 \begin{equation}\label{limit}
\lim_{B\to 0}{eBV\over 2\pi}\delta(\epsilon-{eB\over 2})={L\over \pi\sqrt{2\epsilon}}
\end{equation}
a relation which has to be understood as arising in the thermodynamic limit $\omega\to 0$. 

To arrive at (\ref{limit}), one could as well   consider  directly  the $1$-body 
  harmonic LLL spectrum (\ref{LLLomega}) and  harmonic 1d  spectrum (\ref{harspec})
\begin{equation}\label{limitbis}
  E=(\omega_t-\omega_c)(l_i+1)+\omega_c;
\quad E= \omega(l_i +{1\over 2})
\end{equation}
They are such that the latter is the vanishing $B$ limit of the former, so it is the case for the  corresponding 1-body partition functions. Taking then\footnote{The order of limits is crucial here: first the limit $B\to 0$, then the thermodynamic limit $\omega\to 0$.} the thermodynamic limit $\beta\omega\to 0$, i.e. (\ref{thermolimit}),  implies  the relation  $\lim_{B\to 0}Z_{LLL}= Z_{0}$, where $Z_{LLL}$ is, as above, the LLL partition function and   $Z_0$ is the free  partition function in one dimension. Consequently for the densities of states (the inverse Laplace transforms) the relation (\ref{limit}) follows. This result has its roots in the different  energy gaps of the spectra  (\ref{limitbis}) at small $\omega$: in the harmonic LLL case, the gap  behaves like $\omega^2/(2\omega_c)$, whereas, in the 1d harmonic case, the gap  is  $\omega$.    

The relation (\ref{limit}) could also have been understood from the 1-body eigenstates themselves. In the limit $B\to 0$, the LLL induced harmonic analytic eigenstates are,  from (\ref{LLLomega}), 
\begin{equation}\label{red}
\sqrt{({\omega^{l_i+1}\over \pi l_i! })}z_i^{l_i} e^{-{1\over 2}\omega z_i\bar z_i}
\end{equation}
There is only one parameter $\omega$ left so that the states in (\ref{red}) can be put in one-to-one    correspondence  with the  Hermite polynomials (\ref{harspec}) via the Bargmann transform  
\begin{equation}\label{limitter}
 \sqrt{\omega^{l_i+1}} z_i^{l_i}=\omega\int_{-\infty}^{\infty} dx_i {1\over \sqrt{2^{l_i}}}e^{-\omega(x_i^2-z_ix_i\sqrt 2+z_i^2/2)}H_{l_i}(\sqrt\omega x_i)
\end{equation}
From (\ref{limitter}) one can infer \cite{Brink} that the $N$-body harmonic anyon  eigenstates (\ref{etoui})  are a coherent state representation of the $N$-body harmonic Calogero  eigenstates.

From all these considerations (thermodynamics, eigenstates,...) it follows that the  vanishing magnetic field limit\footnote{Since one has ended up by taking the limit $B\to 0$, one could have   avoided right from the beginning to  introduce  a $B$ field, and started directly from the harmonic $N$-body anyon model. What has been done above by taking the limit $B\to 0$  is nothing but  to  project the harmonic anyon model on the LLL induced harmonic subspace (\ref{red}) (the $B$ field and its LLL should  still  be invoked  to justify the selection of the  LLL quantum numbers in the 2d harmonic basis)  and to recognize that the projected harmonic anyon model is the harmonic Calogero model.  This relation remains  true in the thermodynamic limit  $\omega\to 0$. } of the LLL-anyon model is the Calogero model itself. 
It seems paradoxical  to consider such a limit in  the LLL which assumes a priori a strong magnetic field. Still, doing so, one  has dimensionally reduced the  2d anyon model to the 1d Calogero model. This dimensional reduction    has a simple geometrical interpretation.  
The LLL induced harmonic states (\ref{red}) are localized in the vicinity of  circles of radius $l_i/\omega$. In the thermodynamic limit, one has $l_i\to\infty$ with $l_i\omega=k_i^2/2$ kept fixed. It follows that  the corresponding 1d Hermite polynomials $H_{l_i}$, which become in this limit plane waves of momentum $k_i$, have a radius of localization diverging like $k_i^2/\omega^2$.
The dimensional reduction which has taken place consists in going at infinity on the  edge of the plane: in the thermodynamic limit,  the Calogero model  can be viewed as  the edge projection of the anyon model.

The LLL anyon thermodynamics, or, equivalently, the Haldane/exclusion thermodynamics, and the Calogero thermodynamics as well, have been the subject of an intense activity
since the mid-nineties.  Let us   mention their relevance  in more abstract contexts,  such as  conformal field theories \cite{Schou}.
On the experimental side,  FQHE edge currents can be modelled by quasiparticles with fractional statistics, which in turn might affect their transport  properties such  as  the current shot noise \cite{Ouvry7, Italiens}.

\section{Minimal Difference Partitions and Trees}

Up to now one has been concerned with  quantum mechanical models defined by a microscopic quantum Hamiltonian. Both the LLL anyon and Calogero  models  have been shown to have a thermodynamics controlled by (\ref{thermoquarquar}) and (\ref{thermoquarquin}). Let us leave quantum mechanics and address a pure  combinatorial problem, 
the minimal difference partition  problem \cite{Andrews}. 
Consider the number $\rho(E,N)$ of partitions of an integer $E$ into $N$ integer parts where each part differs from the next by at least an integer $p$ and the smallest part is $\ge l$. Usual integer partitions correspond to $p=0$ and $l=1$, whereas  restricted partitions, where the parts have to be different, correspond to $p=1$ and $ l=1$.   

\begin{figure}[htbp]
\epsfxsize=12cm
\centerline{\epsfbox{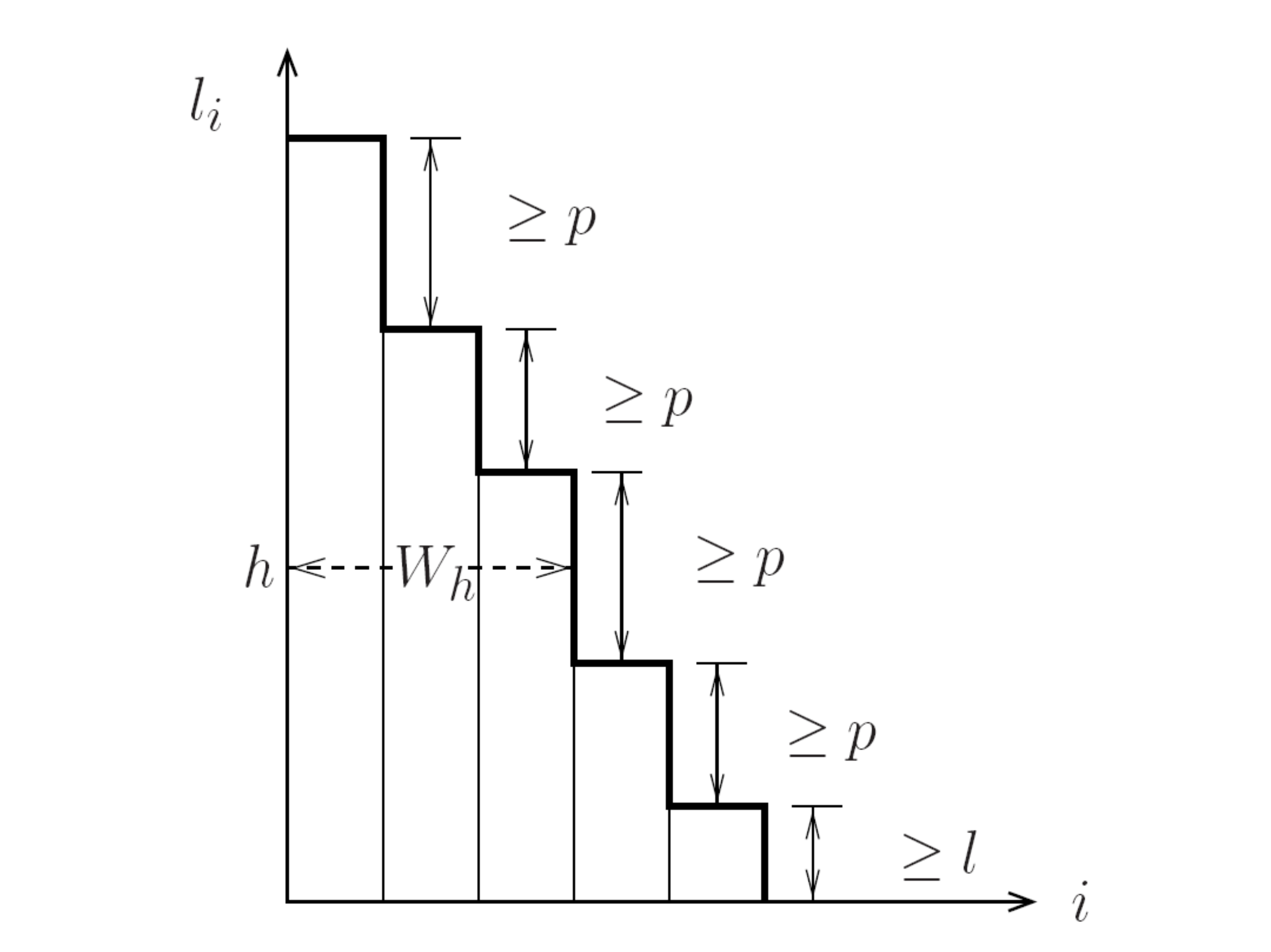}}
\caption{A minimal difference partition configuration, or Young diagram.
The column heights are such that $(l_i-l_{i+1})\ge p$  for
$i=1,2,\ldots ,N-1$ and $l_i\ge l$. Their total
height is $E=\sum_{i=1}^N l_i$. $W_{ h}$ is the width of the Young diagram at height $h$, i.e. the number of columns whose  heights $\ge  h $.}
\label{fig:mdp}
\end{figure}

It is known that
\be\label{ZN} \sum_{E}\rho(E,N)x^E={x^{lN+pN(N-1)/2}\over (1-x)(1-x^2)...(1-x^N)}  \ee
The $\rho(E,N)$ generating function $Z(x,z)=\sum_{E,N}^{\infty}\rho(E,N) x^E z^N$
factorizes when $p=0$ or $p=1$ 
\be\label{fact}  p=0,\quad Z(x,z)=\prod_{i=0}^{\infty}{1\over 1-x^{l+i}z};\quad p=1,\quad
Z(x,z)=\prod_{i=0}^{\infty}{( 1+x^{l+i}z)}\ee
In  terms of bosons or fermions,  (\ref{fact})  is the grand partition function for a bosonic or fermionic gas with fugacity $z$ and, denoting $ x=e^ {-\beta} $, temperature $T=1/ \beta $ where
\be E=\sum_{i=0}^{\infty}n_i(l+i)\quad N=\sum_{i=0}^{\infty}n_i\ee
with $n_i=0,1,2,...$ in the Bose case $(p=0)$ and  $n_i=0,1$ in the Fermi case $(p=1)$.
Equivalently
\be E=\sum_{i=1}^N l_i\ee with  $ l\le l_1\le l_2 \le \ldots \le l_N$ (Bose) or
 $ l\le l_1< l_2 <\ldots < l_N$ (Fermi).

When $p$ is an integer $\ge 2$,  (\ref{ZN})
can be regarded as the $N$-body partition function of an interacting  bosonic gas with the $N$-body  spectrum 
\be\label{spectrum} E=\sum_{i=1}^N l_i+pN(N-1)/2;\quad l\le l_1\le l_2 \le \ldots \le l_N\ee
Clearly, (\ref{spectrum})  goes beyond the Fermi point $p=1$ and describes some kind of ''superfermions''. 
In  contrast to the Bose and Fermi cases, a factorization such as  (\ref{fact})  is not possible, due to  the interacting nature of (\ref{spectrum}). One has instead
the functional relation 
\be Z(x,z)=Z(x,xz)+x^lzZ(x,x^pz)\ee 
which embodies the combinatorial identity
\be\label{comb} \rho(E,N)=\rho_{0}\bigg(E-p{N(N-1)\over 2},N\bigg)\ee
where $ \rho_0(E,N)$ stands for the usual partition counting.

One could push \cite{Maj1} this analysis further to $p$  real positive. When $p\in[0,1]$ and $l=1$, one would obtain a partition problem interpolating between the usual (bosonic) one   and the restricted (fermionic) one. It is manifest that,  if $p$ is replaced by $ -\alpha$, the spectrum (\ref{spectrum})  coincides, under a rescaling and up to an irrelevant zero-point energy, with  the $N$-body quantum spectrum (\ref{energycal}) of the  harmonic Calogero model.
In a partition problem, one is interested in the large $E$ and  $N$  asymptotic behavior of $\rho(E,N)$, which corresponds to the regime $x\to 1$, i.e. $\beta\to 0$. Consider  the cluster expansion $\ln Z(x,z)=\sum_{n=1}^{\infty}b_nz^n$. In the limit $\beta\to 0$ 
  one obtains
\be b_{n}={1\over \beta}{e^{-nl\beta}\over n^2} \prod_{k=1}^{n-1}(1-{p n\over k});\quad b_1={1\over \beta}e^{-l\beta}\label{cluspar}\ee
The limit $\beta\to 0$ should not be confused  with the thermodynamic limit in quantum systems. There is no thermodynamic limit prescription like (\ref{thermolimit}). Still, using (\ref{thermoquarbis}) (with $\omega_c$  replaced by $l$) and  taking care of the unwanted $1/n$ factor in (\ref{cluspar}), one obtains, provided that
$ze^{-\beta l}<1$,
\be\label{usbis} \ln Z(\beta,z)=\int_l^{\infty} \ln y(ze^{-\beta\epsilon})d\epsilon\ee
with
\be \label{us} y-ze^{-\beta\epsilon}y^{1-p}=1\ee
This is again of the form (\ref{thermoquarquar}) and (\ref{thermoquarquin}), the statistical parameter $-\alpha$ being replaced by the minimal difference partition parameter $p$, and  the 1-body density of states being the Heaviside function  $\rho(\epsilon)=\theta(\epsilon-l)$. The minimal difference partition combinatorics is equivalently described, in the small $\beta$ limit, by a gas of particles obeying   exclusion statistics with a uniform density of states\footnote{There is no microscopic quantum Hamiltonian leading to (\ref{usbis}) and (\ref{us}).}. 

This correspondence happens to be useful technically:  (\ref{usbis})  and (\ref{us})  are the  building blocks of the minimal difference partition asymptotics.
The average  integer  $\bar E= -{\partial \ln Z(\beta,z)/ \partial \beta}=\int_l^{\infty} n\epsilon d\epsilon $ and the average number of  integer parts  $\bar N= z {\partial \ln Z(\beta,z)/ \partial z}=\int_l^{\infty} n d\epsilon$, are both given in terms of $n =z\partial \ln y/\partial z$, the mean occupation number at "part" $\epsilon$ and fugacity $z$, which satisfies 
\be ze^{-\beta\epsilon}={n\over (1+(1-p)n)^{1-p}(1-pn)^p}
\quad {\rm with}\quad n\le {1\over p}\ee
One obtains
\be\label{simple} \bar E-l\bar N={1\over \beta}\ln Z(\beta,z) \quad{\rm }\quad\quad \bar N={1\over \beta}\ln y(ze^{-\beta l})\ee
so that the entropy\footnote{The simple expression in (\ref{simple}) for $\bar N$ is possible because of the constant density of states.} $S\equiv \ln Z(\beta,z)+\beta\bar E-(\ln z) \bar N$
rewrites as 
\be\label{entropy} S= 2\beta\bigg(\bar E-l\bar N-{p\over 2}{\bar N}^2\bigg)-\bar N\ln(1-e^{-\beta\bar N})\ee
with
\be\label{energyprime}  \bar E-l\bar N-{p\over 2}\bar N^2=-{1\over \beta^2}\int_0^{1-e^{-\beta\bar N}}{\ln(1-u)\over u}du\ee
Inverting (\ref{energyprime}) gives $\beta$ as a function of $\bar E$ and $\bar N$ so that  the entropy $S$ in (\ref{entropy}) becomes a function of $\bar E$ and $\bar N$ only.
Doing so, one has  a definite information \cite{Maj1} on the asymptotic behavior of $\rho( E, N)\simeq e^{S( E, N)}$ when $E$ and $N$ are large, and also, of $\rho( E)=\sum_{N=1}^{\infty}\rho( E, N)$  when $E$ is large. One obtains  a generalization of the Hardy-Ramajunan  asymptotics \cite{Hardy} to the minimal difference  partition problem.
One can also obtain \cite{Maj2} the average limit shape of the Young diagrams associated with the minimal difference partition problem, generalizing the usual partition limit shape \cite{Vershik}. The limit shape at a part of height $ h$ depends  solely on the statistical function $y$  evaluated at $\epsilon= h$ and at $z=1$
\be \beta {\bar W}_h=\ln y(e^{-\beta h})
\ee 
where $\beta $ scales as  $\beta^2 E=\int_0^{\infty}\ln y(e^{-\epsilon})d\epsilon$.

So far $p$ being a positive integer has insured  that the $N$-body spectrum in (\ref{spectrum}) is well defined.
However,  $y$ in (\ref{us}) is still meaningful when  $p$ is a negative integer. It is the $(1-p)$-ary tree generating function, so that the coefficient at order $n$ of its expansion in powers of $ze^{-\beta\epsilon}$  as given in (\ref{thermoquinbisbisbis}) (with $-\alpha$ replaced by $p$) is the number of ways to build a $(1-p)$-ary tree with $n$ nodes. For example, at $p=-1$, $y$ generates the Catalan numbers associated with binary trees.

Consider, as a toy model \cite{Maj3},  the  factorized $(1-p)$-ary tree   generating function 
\be\label{factbis}  Z(x,z)=\prod_{i=0}^{\infty}{y(zx^{l+i})}\ee
where  $y$ satisfies  (\ref{factbis}) with $\epsilon=l+i$. (\ref{factbis}) narrows down to (\ref{fact}) when $p=0$ (Bose case).
Its combinatorial interpretation is that  $\rho(E,N)$  deduced from  (\ref{factbis})  counts the number of  
usual partitions of an integer  $E$ into $N$  integer 
parts bigger or equal to $l$, with an additional  degeneracy stemming from the $(1-p)$-tree arborescence when, in  a given partition, a part
occurs $n$ times. This enlarged  degeneracy  goes beyond the Bose point to define some kind of  "superbosons".

One can  analytically continue  $p$ to the whole negative real axis. In the large $E$ and $N$ limit, i.e. $\beta$ smaller and smaller,  one encounters a maximal  temperature  beyond which it is not possible to heat the system. Indeed, from (\ref{us}) it follows that  $y(zx^{l+i})$ in (\ref{factbis})  obeys    to 
$y-ze^{-\beta(l+i)}y^{1-p}=1$, which is well defined  only if \cite{Bytsko} 
\be ze^{-\beta l}<(1-p)^{p-1}(-p)^{-p}<1\ee
When $z=1$, it   defines  a dimensionless "Hagedorn temperature" 
\be T={l\over (1-p)\ln(1-p)+p\ln(-p)}\ee
 just below  which
$E$ and $N$ become large so that the asymptotic of $\rho(E,N)$ can be addressed. 

\section{Conclusion}
In two dimensions intermediate anyonic statistics interpolating from Bose to Fermi statistics are allowed. Their definition does not involve anything else than the usual  concept at the basis of quantum statistics, namely free particles endowed with particular boundary exchange conditions on their $N$-body wavefunctions. It happens that these boundary conditions have a much richer structure in two dimensions than in three and higher dimensions. This  in turn can be understood in terms of the topology of paths in  the $N$-particle configuration space, where non trivial braiding occurs in two dimensions, and not in higher dimensions.  A flux-charge composite picture emerges to encode the braiding statistics in physical terms, via topological Aharonov-Bohm interactions  and singular magnetic fields.

The anyon model as such is certainly  fascinating as far as quantum mechanics is concerned, but it remains an abstract construction whose  complexity is daunting. However, when projected onto the LLL of an external magnetic field, the model becomes tractable and, even more, solvable. The LLL set up is clearly adapted to the QHE and to the FQHE physics. Haldane/exclusion statistics, which can be obtained  as a LLL-anyon mean-field picture  in the screening regime,    leads to  LLL-anyon thermodynamics.
 
It would certainly be  rewarding if  LLL anyons could  be relevant experimentally, for example by uncovering some experimental  hints  at FQHE filling factors of the existence  of quasiparticles with anyonic/exclusion statistics. Fractional charges have already been seen in shot noise FQHE experiments \cite{Glattli},  but the nontrivial statistical nature of the charge carriers in FQHE edge currents has so far remained elusive in experiments which rely mainly  on  Aharonov-Bohm interferometry \cite{Camino}.
Note also a recent proposal for the possible experimental tracking of  abelian and nonabelian anyonic statistics in Mach-Zehnder interferometers \cite{Kitaev2}.

Finally, on the theoretical side, physical  interactions,  together with topological anyonic interactions,  should also be taken into account in order to  produce  more realistic models.

\vspace{1cm}

{\bf Acknowledgements}

\vspace{1cm}

I would like to thank  Alain Comtet and Stefan Mashkevich for past and present collaborations, and for correcting and improving  the text. My thanks also to Tobias Paul and Sanjib Sabhapandit for helping me with  the Figures.


\begin{thebibliography}{10}


\bibitem{LM} J.M.~Leinaas, \JM,
On the theory of identical particles,
Nuovo Cimento 37B (1977) 1--23. For an earlier work on the subject see: M.G.G. Laidlaw,  
C.M. de Witt,
Feynman Functional Integrals for Systems of Indistinguishable Particles, Phys. Rev D3  (1971) 1375. 

\bibitem{Forte}  For a recent review on spin  issues related to  statistics see:  S. Forte, 
  Spin in quantum field theory,
Proceedings of the 43th Internationale Univeritastswochen fur Theoretiche Physik, Schladming, Austria (2005). arXiv: hep-th/0507291

\bibitem{Nonab}
G. Moore,  N. Seiberg, Polynomial equations for rational
conformal field theories, Phys. Lett. B 212 (1988) 451.
Classical and quantum conformal
field theory, Commun. Math. Phys. 123 (1989) 177.
G. Moore,   N. Read, Nonabelions in the fractional quantum
Hall effect, Nucl. Phys. B 360(2-3) (1991) 362.




\bibitem{Comp} A.Yu.~Kitaev,
 Fault-tolerant quantum computation by anyons,
Ann.~Phys. 303  (2003) 2--30.



\bibitem{W-82-0} F.~Wilczek,
 Magnetic flux, angular momentum, and statistics,
Phys.~Rev.~Lett. 48 (1982) 1144--1146.
 Quantum mechanics of fractional-spin particles,
Phys.~Rev.~Lett. 49 (1982) 957--959.

\bibitem{AB} Y.~Aharonov, D.~Bohm,
 Significance of electromagnetic potentials in quantum theory
Phys. Rev. 115 (1959) 485--491.
For an earlier work on the subject see: W.~Ehrenberg, R.W.~Siday,
 The Refractive index in electron optics and the principles of dynamics,
Proc. Phys. Soc. London B62 (1949) 8--21.


\bibitem{Cham}  R.G. Chambers, 	
Shift of an electron interference pattern by enclosed magnetic Flux, 
Phys.~Rev.~Lett. 5 (1960) 3.


\bibitem{Tono} Tsuyoshi Matsuda, Shuji Hasegawa, Masukazu Igarashi, Toshio Kobayashi, Masayoshi Naito, Hiroshi Kajiyama, Junji Endo, Nobuyuki Osakabe, Akira Tonomura,  Ryozo Aoki,   Magnetic field observation of a single flux quantum by electron-holographic interferometry,
Phys. Rev. Lett. 62 (1989) 2519. 


\bibitem{Chern} W.~Siegel,
Unextended superfields in extended supersymmetry,
Nucl.~Phys. B156 (1979) 135--143.
J.F.~Schonfeld,
A mass term for three-dimensional gauge fields,
Nucl.~Phys. B185 (1981) 157--171.
R. Jackiw, S. Templeton,
 How super-renormalizable interactions cure their infrared divergences,
Phys.~Rev. D23 (1981) 2291--2304.
 S.~Deser, R.~Jackiw, S.~Templeton,
 Three-dimensional massive gauge theories,
Phys.~Rev.~Lett. 48 (1982) 975--978.
 Topologically massive gauge theories,
Ann. Phys. (N.Y.) 140  (1982) 372--411.

\bibitem{Pert1} J.~McCabe, S.~Ouvry,
 Perturbative three-body spectrum and the third virial coefficient in the anyon model,
Phys.~Lett. B260 (1991) 113--119.
\bibitem{Pert2} \SO, $\delta$ perturbative interactions in the Aharonov-Bohm and anyon models,
Phys.~Rev. D50 (1994) 5296--5299.
A.~Comtet, \SM, \SO, Magnetic moment and perturbation theory with singular magnetic fields,
Phys.~Rev. D52 (1995) 2594--2597.

\bibitem{selfadjoint} C.~Manuel, R.~Tarrach,
Contact interactions of anyons,
Phys.~Lett. B268 (1991) 222--226.

\bibitem{Myr2} For a review on the anyon model, see (among others):  J. Myrheim, Anyons, Les Houches LXIX Summer School {\it ''Topological aspects of low dimensional systems''} (1998) 265--414.


\bibitem{Wu1} Y.-S.~Wu,
 Multiparticle quantum mechanics obeying fractional statistics,
Phys.~Rev.~Lett. 53 (1984) 111--114.

\bibitem{N-body} A.P.~Polychronakos,
 Exact anyonic states for a general quadratic hamiltonian,
Phys.~Lett. B264 (1991) 362--366.
 C.~Chou,
 Multianyon spectra and wave functions,
Phys.~Rev. D44 (1991) 2533--2547.
 \SM,
 Exact solutions of the many-anyon problem,
Int.~J.~Mod.~Phys. A7 (1992) 7931--7942.
G.~Dunne, A.~Lerda, S.~Sciuto, C.A.~Trugenberger,
 Exact multi-anyon wave functions in a magnetic field,
Nucl.~Phys. B370 (1992) 601--635.
 A.~Karlhede, E.~Westerberg,
Anyons in a magnetic field,
Int.~J.~Mod.~Phys. B6 (1992) 1595--1621.
 \SM,
 Towards the exact spectrum of the three-anyon problem,
Phys.~Lett. B295 (1992) 233--236.


\bibitem{Second} D.~Arovas, R.~Schrieffer, F.~Wilczek, A.~Zee,
 Statistical mechanics of anyons,
Nucl.~Phys. B251 (1985) 117--126.

\bibitem{Secondbis}
A.~Comtet, Y.~Georgelin, S.~Ouvry,
Statistical aspects of the anyon model,
J. Phys. A: Math. Gen. 22 (1989) 3917--3926.


\bibitem{Third} D.~Sen,
Spectrum of three anyons in a harmonic potential and the third virial
coefficient,
Phys.~Rev.~Lett. 68 (1992) 2977--2980.
M.~Sporre, J.J.M.~Verbaarschot, I.~Zahed,
Anyon spectra and the third virial coefficient,
Nucl.~Phys. B389 (1993) 645--665.


\bibitem{Num} M.~Sporre, J.J.M.~Verbaarschot, I.~Zahed,
Numerical solution of the three-anyon problem,
Phys.~Rev.~Lett. 67 (1991) 1813--1816.
M.V.N.~Murthy, J.~Law, M.~Brack, R.K.~Bhaduri, Quantum spectrum of three anyons in an oscillator potential,
Phys.~Rev.~Lett. 67 (1991) 1817--1820.
M.~Sporre, J.J.M.~Verbaarschot, I.~Zahed,
Four anyons in a harmonic well,
Phys.~Rev. B46 (1992) 5738--5741.

\bibitem{Semi} R.K.~Bhaduri, R.S.~Bhalerao, A.~Khare, J.~Law, M.V.N.~Murthy,
 Semiclassical two- and three-anyon partition functions,
Phys.~Rev.~Lett. 66 (1991) 523--526.
 F. Illuminati, F. Ravndal, J.Aa. Ruud,
 A semi-classical approximation to the three-anyon spectrum,
Phys.~Lett. A161 (1992) 323--325.
J.Aa.~Ruud, F.~Ravndal,
Systematics of the ${N}$-anyon spectrum,
Phys.~Lett. B291 (1992) 137--141.


\bibitem{Ouvry1} A.~Comtet, J.~McCabe, S.~Ouvry,
Perturbative equation of state for a gas of anyons,
Phys.~Lett. B260 (1991) 372--376.

\bibitem{Ouvry2} A.~Dasni\`eres de Veigy, S.~Ouvry,
Perturbative equation of state for a gas of anyons: Second order,
Phys. Lett. B291 (1992) 130--136.
 Perturbative anyon gas,
Nucl. Phys. B388 (1992) 715--755.



\bibitem{Myr1} \JM, \KO,
 The third virial coefficient of free anyons, Phys. Lett. B299 (1993) 267--272.
 \SM, \JM, \KO, The third virial coefficient of  anyons revisited,
Phys. Lett. B382 (1996) 124--130.
 A.~Kristoffersen, \SM, \JM, \KO,
 The fourth virial coefficient of anyons,
Int.~J.~Mod.~Phys. A 11 (1998) 3723--3747.
\SM, \JM, \KO, R.~Rietman,
 The nature of the three-anyon wave functions,
Phys.~Lett. B348 (1995) 473--480.




\bibitem{Laughlin} R.B.~Laughlin,
 Quantized Hall conductivity in two dimensions,
Phys.~Rev. B23 (1981) 5632--5633.
Anomalous quantum Hall effect:
An incompressible quantum fluid with fractionally charged excitations,
Phys.~Rev.~Lett. 50 (1983) 1395--1398.
Quantized motion of three two-dimensional electrons in a strong magnetic field,
Phys.~Rev. B27 (1983) 3383--3389.  See also: F.D.M.~Haldane,
Fractional quantization of the Hall effect:
A hierarchy of incompressible quantum fluid States,
Phys.~Rev.~Lett. 51 (1983) 605--608.


\bibitem{FQHE}  K.~von Klitzing, G.~Dorda, M.~Pepper,
New method for high-accuracy determination
of the fine-structure constant based on quantized Hall resistance,
Phys.~Rev.~Lett. 45 (1980) 494--497.
D.C.~Tsui, H.L.~St\"ormer, A.C.~Gossard,
 Zero-resistance state of two-dimensional electrons
in a quantizing magnetic field,
Phys.~Rev. B25 (1982) 1405--1407.
M.A.~Paalanen, D.C.~Tsui, A.C.~Gossard,
 Quantized Hall effect at low temperatures,
Phys.~Rev. B25 (1982) 5566--5569. H.L.~St\"ormer, A.~Chang, D.C.~Tsui, J.C.M.~Hwang,
A.C.~Gossard, W.~Wiegmann,
 Fractional quantization of the Hall effect,
Phys.~Rev.~Lett. 50 (1983) 1953--1956.

\bibitem{FQHEbis} D.P.~Arovas, R.~Schrieffer, F.~Wilczek,
 Fractional statistics and the quantum Hall effect,
Phys. Rev. Lett. 53 (1994) 722--725.
 B.I.~Halperin,
 Statistics of quasiparticles and the hierarchy of fractional quantized Hall states,
Phys.~Rev.~Lett. 52 (1984) 1583--1586.

\bibitem{Berry} H.~Kj{\o}nsberg, \JM,
Numerical study of charge and statistics of Laughlin quasiparticles,
Int. J. Mod. Phys. A14 (1999) 537--557.
 D.~Banerjee,
Topological aspects of phases in fractional quantum Hall effect,
Phys.~Lett. A269 (2000) 138--143.




\bibitem{Camino} F.E.~Camino, W.~Zhou, V.J.~Goldman,
Aharonov-Bohm electron interferometer in the integer quantum Hall regime, 
arXiv: cond-mat/0503456.
     Experimental realization of Laughlin quasiparticle interferometers,
    Proc. of EP2DS-17 (Genoa, Italy, 2007). arXiv:0710.1633.


\bibitem{Ouvry3}  A.~Dasni\`eres de Veigy, \SO,
 Equation of state of an anyon gas in a strong magnetic field,
Phys.~Rev.~Lett. 72 (1994) 600--603.



\bibitem{Fermi} E. Fermi was the first to introduce an harmonic well confinement to compute thermodynamical quantities:  E. Fermi, Sulla quantizzazione del gas perfetto monoatomico, Rend. Lincei 3 (1926) 145. In the anyon  context,  the harmonic well confinement was first used in \cite{Secondbis}. See also: \KO,
On the harmonic oscillator regularization of
partition functions,
Trondheim preprint No.~13 (1992). 


\bibitem{Masheigen}\SM, \JM, \KO, R.~Rietman,
 Anyon trajectories and the systematics of the three-anyon spectrum,
Int.~J.~Mod.~Phys. A 11 (1996) 1299--1313.


\bibitem{Ouvry4}  S. Ouvry, On the relation between the anyon and the Calogero Models, Phys. Lett. B510 (2001) 335.

\bibitem{Ouvry4bis} S. Isakov, G. Lozano, S. Ouvry, Non abelian Chern-Simons
particles in an external magnetic field,
     Nucl. Phys. B552 [FS] (1999) 677.

\bibitem{Haldane} F.D.M.~Haldane,
``Fractional statistics'' in arbitrary dimensions:
A generalization of the Pauli principle,
Phys.~Rev.~Lett. 67 (1991) 937--940.


\bibitem{Anti}\SM, \SO,
 The lowest Landau level anyon equation of state
in the anti-screening regime,
Phys.~Lett. A310 (2003) 85--94.

\bibitem{Bose}  
N.K. Wilkin, J.M. Gunn, R.A. Smith, Do attractive bosons condense?,
Phys. Rev. Lett. 80 (1998) 2265.




\bibitem{Ouvry5} A.~Dasni\`eres de Veigy, S.~Ouvry,
One-dimensional statistical mechanics for 
identical particles : the Calogero and anyon cases, 
Mod. Phys. Lett. B 9 (1995) 271.

\bibitem{Poly} 
A.P.~Polychronakos, Probabilities and path-integral realization of exclusion statistics. arXiv: hep-th/9503077.
See also: Generalized statistics in one dimension, Les Houches LXIX Summer School {\it ''Topological aspects of low dimensional systems''} (1998) 415--472.

\bibitem{Bytsko} A.G. Bytsko,
 Haldane-Wu statistics and Rogers dilogarithm,
 Zap. Nauchn. Semin. POMI 291 (2002) 64-77, J.Math.Sciences 125 (2005) 136-143.

\bibitem{Wu2} Y.S.~Wu,
 Statistical distribution for generalized ideal gas
of fractional-statistics particles,
Phys.~Rev.~Lett. 73 (1994) 922--925.

\bibitem{Wilczek2}  C. Nayak, F.~Wilczek,
Exclusion statistics: Low-temperature properties, fluctuations, duality, and applications,
Phys. Rev. Lett. 73 (1994) 2740. S. Chaturvedi, V. Srinivasan, Microscopic interpretation of Haldane's semion statistics, Phys. Rev. Lett. 78 (1997) 4316. M.V.N. Murthy, R. Shankar, Exclusion statistics: A resolution of the problem of negative weights, Phys. Rev. B 60 (1999) 6517.

\bibitem{Italiens} G. Gomila, L. Reggiani, Fractional exclusion statistics and shot noise in ballistic conductors (2001), Phys. Rev. B 63 (2001) 165404.

\bibitem{Calogero} F. Calogero,
 Solution of a three-body problem in one dimension,
J. Math. Phys. 10 (1969) 2191--2196.
 Ground state of a one-dimensional $N$-body system,
J. Math. Phys. 10 (1969) 2197--2200.
 Solution of the one-dimensional $N$-body problems with quadratic and/or inversely quadratic pair potentials,
J. Math. Phys. 12 (1971) 419--436.


\bibitem{Calogero1} For a recent review on the Calogero model, see (among others): A.P. Polychronakos, Physics and mathematics of Calogero particles. arXiv: 
hep-th/0607033.

\bibitem{Isakov} A.P. Polychronakos, Non-relativistic bosonization and fractional statistics, Nucl. Phys. {\bf B324} (1989) 597.
Exchange operator formalism for integrable systems of particles, Phys. Rev. Lett.
{\bf 69} (1992) 703.
S. B. Isakov, Fractional statistics in one dimension: Modeling by means of $1/x^2$ interaction and statistical mechanics, Int. J. Mod. Phys. A9 (1994) 2563. Generalization of statistics for several species of identical particles, Mod. Phys. Lett. B8 (1994) 319.
 Bosonic and fermionic single-particle states
in the Haldane approach to statistics for identical particles,
Phys.~Rev. B53 (1996) 6585--6590.   D. Bernard, Y.-S. Wu,  A Note on statistical interactions and the thermodynamic Bethe ansatz,
(1994). arXiv: cond-mat/9404025. [



\bibitem{Sutherland} B. Sutherland,
 Quantum many-body problem in one dimension: Ground state,
J. Math. Phys. 12 (1971) 246--250.
 Quantum many-body problem in one dimension: Thermodynamics,
J. Math. Phys. 12 (1971) 251--256.




\bibitem{Brink}
L. Brink, T.H. Hansson, S. Konstein, M.A. Vasiliev,
The Calogero model  anyonic representation, fermionic extension and supersymmetry,
Nucl. Phys. B, Volume 401, Issue 3 (1993)  591-612.

\bibitem{Schou} K.~Schoutens,
Exclusion statistics in conformal field theory spectra,
Phys.~Rev.~Lett. 79 (1997) 2608--2611. P. Fendley, K. Schoutens, Cooper pairs and exclusion statistics from coupled free-fermion chains,
 J. Stat. Mech. 0207 (2007) 17.

\bibitem{Ouvry7} S. Isakov, T. Martin, S. Ouvry,  Conductance and shot noise for particles with exclusion
statistics,  Phys. Rev. Lett. 83 (1999) 580.


\bibitem{Andrews} G.E. Andrews, {\it The Theory of partitions}, Cambridge University
Press, Cambridge (1998). G.E. Andrews, R. Askey, R. Roy, {\it Special functions}, 
Encyclopedia of Mathematics and its 
applications 71, Cambridge University Press, Cambridge (1999).


\bibitem{Maj1} A. Comtet, S.N. Majumdar, S. Ouvry, Integer partitions and exclusion statistics, J. Phys. A: Math. Theor. 40  (2007) 11255.

\bibitem{Hardy} G.H. Hardy, S. Ramanujan, Proc. London. Math. Soc. {17} (1918) 75.

\bibitem{Maj2} A. Comtet, S.N. Majumdar, S. Ouvry, S.  Sabhapandit, Integer partitions and exclusion statistics: Limit shapes and the largest part of Young diagrams, J. Stat. Mech.  (2007) P10001.  

\bibitem{Vershik} A.M. Vershik, Statistical mechanics of combinatorial partitions and their limit shapes,  {\it Functional Analysis and Its Applications 30}  (1996) 90.

\bibitem{Maj3}  A. Comtet, S.N. Majumdar, S. Ouvry, S.  Sabhapandit, in preparation.

\bibitem{Glattli}  L. Saminadayar, D.C. Glattli, Y. Jin, B. Etienne, 	
Observation of the e/3 fractionally charged Laughlin quasiparticle, Phys. Rev. Lett 79 (1997) 162.
R. de-Picciotto et al, Nature 389 (1997) 162.


\bibitem{Kitaev2} D. E. Feldman, Y. Gefen, A.Yu. Kitaev, K. T. Law, A. Stern,
Shot noise in anyonic Mach-Zehnder interferometer,
Phys. Rev. B 76 (2007) 085333.


\end{thebibliography}
\end{document}